\documentclass[journal]{IEEEtran}
\usepackage{array}
\usepackage{booktabs}
\usepackage{amsfonts}
\usepackage{amsmath,graphicx,setspace,algorithm,algorithmic,multirow,cases,amssymb,subfigure}
\usepackage{lettrine}
\usepackage{multirow}
\usepackage{xcolor}
\usepackage{times}
\usepackage{graphicx}
\usepackage{amsmath}
\usepackage{indentfirst}
\usepackage{booktabs}
\usepackage{blindtext}
\usepackage{array}
\usepackage{tabularx,booktabs}
\usepackage{color, soul}
\usepackage{bm}
\usepackage{cite}
\usepackage{rotating}
\usepackage{amssymb,mathrsfs,amsmath}
\usepackage[
linkcolor=red,anchorcolor=blue,citecolor=green]{hyperref}
\usepackage{threeparttable}
\usepackage{float}
\newcommand{\RNum}[1]{\uppercase\expandafter{\romannumeral #1\relax}}

\begin{document}

\title{Semantics-Consistent Representation Learning for Remote Sensing Image-Voice Retrieval}

\author{
Hailong~Ning,~\IEEEmembership{}Bin~Zhao,~\IEEEmembership{}and Yuan~Yuan,~\IEEEmembership{Senior Member,~IEEE}

\thanks{Manuscript received XX XX, XXXX.}

\thanks{
This work was supported in part by the National Key R\&D Program of China under Grant 2020YFB2103902, in part by the National Science Fund for Distinguished Young Scholars under Grant 61825603, in part by the Key Program of National Natural Science Foundation of China under Grant 61632018, and in part by China Postdoctoral Science Foundation under Grant 2020TQ0236. \emph{(Corresponding author: Yuan Yuan)}

H. Ning is with Shaanxi Key Laboratory of Ocean Optics, Xi'an Institute of Optics and Precision Mechanics, Chinese Academy of Sciences, Xi'an 710119, P.R. China, the University of Chinese Academy of Sciences, Beijing 100049, P. R. China, and School of Artificial Intelligence, Optics and Electronics (iOPEN), Northwestern Polytechnical University, Xi¡¯an 710072, P.R. China.

B. Zhao and Y. Yuan are with School of Artificial Intelligence, Optics and Electronics (iOPEN), Northwestern Polytechnical University, Xi¡¯an 710072, P.R. China.
}}
\maketitle

\begin{abstract}
With the development of earth observation technology, massive amounts of remote sensing (RS) images are acquired. To find useful information from these images, cross-modal RS image-voice retrieval provides a new insight. This paper aims to study the task of RS image-voice retrieval so as to search effective information from massive amounts of RS data. Existing methods for RS image-voice retrieval rely primarily on the pairwise relationship to narrow the heterogeneous semantic gap between images and voices. However, apart from the pairwise relationship included in the datasets, the intra-modality and non-paired inter-modality relationships should also be taken into account simultaneously, since the semantic consistency among non-paired representations plays an important role in the RS image-voice retrieval task. Inspired by this, a semantics-consistent representation learning (SCRL) method is proposed for RS image-voice retrieval. The main novelty is that the proposed method takes the pairwise, intra-modality, and non-paired inter-modality relationships into account simultaneously, thereby improving the semantic consistency of the learned representations for the RS image-voice retrieval. The proposed SCRL method consists of two main steps: 1) semantics encoding and 2) semantics-consistent representation learning. Firstly, an image encoding network is adopted to extract high-level image features with a transfer learning strategy, and a voice encoding network with dilated convolution is devised to obtain high-level voice features. Secondly, a consistent representation space is conducted by modeling the three kinds of relationships to narrow the heterogeneous semantic gap and learn semantics-consistent representations across two modalities. Extensive experimental results on three challenging RS image-voice datasets, including Sydney, UCM and RSICD image-voice datasets, show the effectiveness of the proposed method.
\end{abstract}


\IEEEpeerreviewmaketitle

\section{Introduction}
\IEEEPARstart{W}{ith} the rapid development of remote sensing (RS) technology, tons of remote sensing images have been produced \cite{8986556, 8385104, zhang2019hierarchical}. There is no doubt that mining useful information from these RS images \cite{8067633} is very important. However, considering that the RS data is too large, it is unpractical to find the useful information manually due to the time-consuming workload. Driven by the practical demand, many researchers pay attention to the large-scale RS data retrieval task \cite{demir2016hashing-based, 8385104, 8067633, Chen2020Supervised}. It can automatically refine precise semantic information contained in the RS data, and has wide application prospects in military targets detection, military audio intelligence generation, and disaster monitoring scenarios \cite{li2018large-scale, chaudhuri2020cmir-net, shi2017can}.

\begin{figure}
\begin{center}
\includegraphics[width=\linewidth]{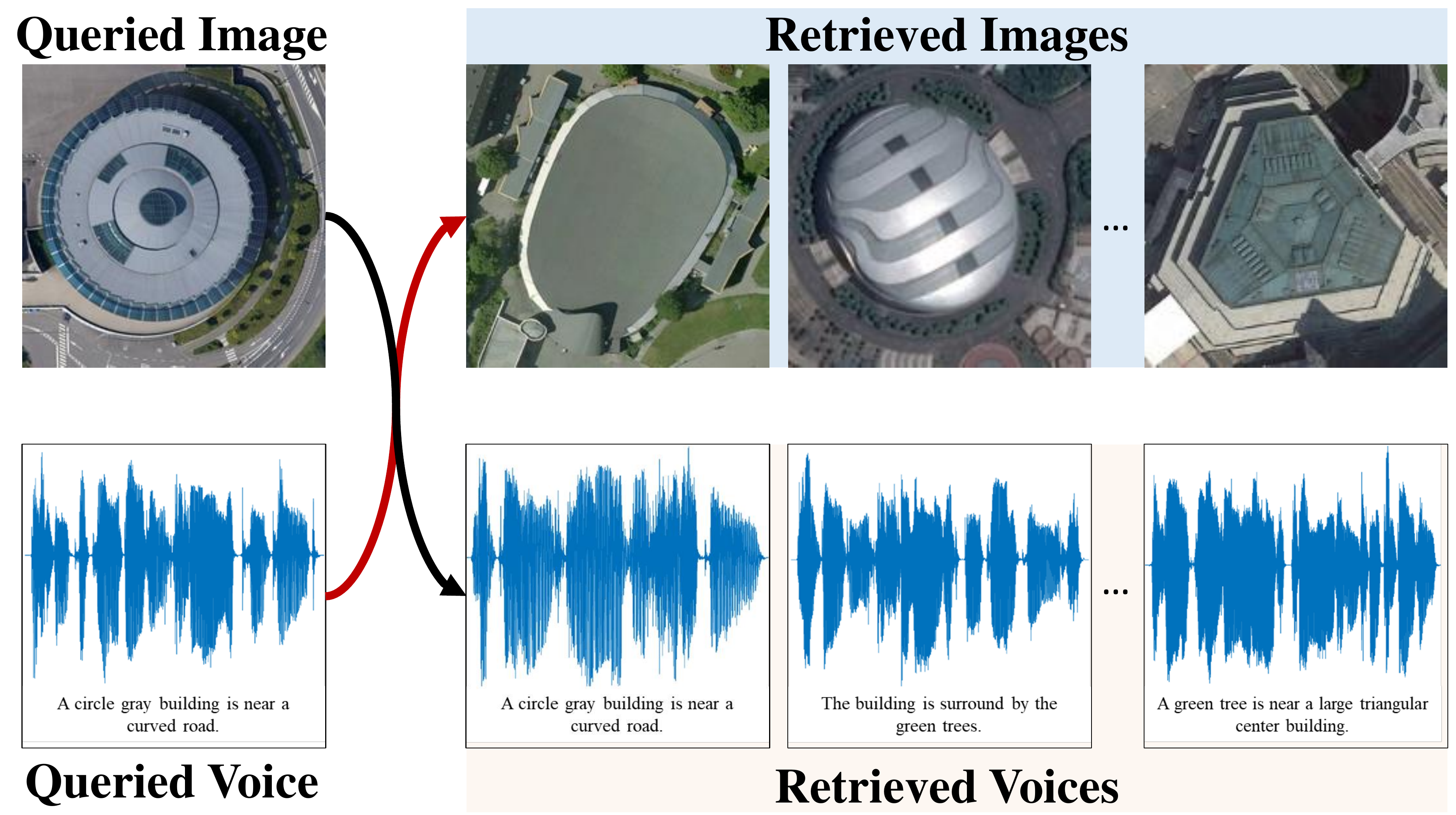}
\renewcommand{\figurename}{Fig.}
\end{center}
    \caption{\small{The schematic diagram of the RS image-voice retrieval task. The goal of the task is to leverage the queried RS image (voice) to retrieve the relevant voices (images).}}
\label{fig:1}
\end{figure}

Generally, the RS data retrieval methods can be roughly divided into two categories: uni-modal retrieval and cross-modal retrieval. Specifically,
the uni-modal RS data retrieval methods conduct the similarity search in the same modality \cite{8954885, li2018large-scale}.For instance, Li {\it et al.} \cite{li2018large-scale} present a large-scale RS image retrieval method to retrieve the RS images with similar semantics. By comparison, the source and target data are from different modalities in cross-modal RS data retrieval methods \cite{8486338, chaudhuri2020cmir-net}. For instance, Chaudhuri {\it et al.} \cite{chaudhuri2020cmir-net} propose a deep neural network to learn a discriminative shared feature space for the input RS images and the corresponding voice based annotations. This paper focuses on the cross-modal RS data retrieval, which is more challenging than uni-modal RS data retrieval due to the heterogeneous semantic gap of cross-modal data.

As shown in Fig. \ref{fig:1}, the task of RS image-voice retrieval is to find the relevant RS image (voice) given a queried voice (image).
Notice that the task is more convenient for humans than the way by keyboard inputting text to retrieve similar RS images, because it is based on human speeches to retrieve target RS images \cite{8896932, 8486338}. As a consequence, the task is more practical in some scenarios that are emergent and not convenient for keyboard input, such as military targets detection, military audio intelligence generation, and disaster monitoring \cite{8486338, shi2017can}.

Due to the practicability of the RS image-voice retrieval task, several related works have been developed \cite{8486338, 8896932, chaudhuri2020cmir-net, chen2019a, 9044618}. These works can be classified into two categories: CNN feature based methods and hash based methods. The former methods conduct the retrieval task by learning CNN features and their consistency relationship \cite{shao2020multilabel}. As an example, Guo {\it et al.} \cite{8486338} propose a deep visual-audio network to learn the correspondence between RS images and voices. The latter methods encode high-dimensional data points into compact binary code for retrieval \cite{9046296}. For instance, Chen {\it et al.} \cite{chen2019a, 9044618} present to integrate the feature learning and hash code learning into a unified framework to achieve fast retrieval property between RS images and voices.

Practically, these works commonly narrow the distance between representations from the paired multi-modalities directly, resulting the similar paired representations for cross-modal retrieval. However, the obtained similar representations by these works are difficult to effectively alleviate the pervasive heterogeneous semantic gap, because the cross-modal retrieval aims to retrieve all the data with the same semantic concept rather than only one data coming from the same pair. As a consequence, it is necessary to fully mine the relationships of both paired and non-paired data within and across modalities. Naturally, we propose to simultaneously model the semantic relationships, including pairwise, intra-modality, and non-paired inter-modality relationships. Specifically, the pairwise relationship models the consistency between the paired image and voice. The intra-modality relationship is to enhance the consistency between two representations from the same modality and belonging to the same semantic concept, thereby avoiding the adverse impact caused by the problems of inter-class similarity and intra-class diversity. The non-paired inter-modality relationship is to enhance the consistency between two non-paired representations from different modalities but belonging to the same semantic concept. In order to further improve the retrieval performance, the pairwise, intra-modality, and non-paired inter-modality relationships should be considered comprehensively.

\begin{figure*}
\begin{center}
\includegraphics[width=\linewidth]{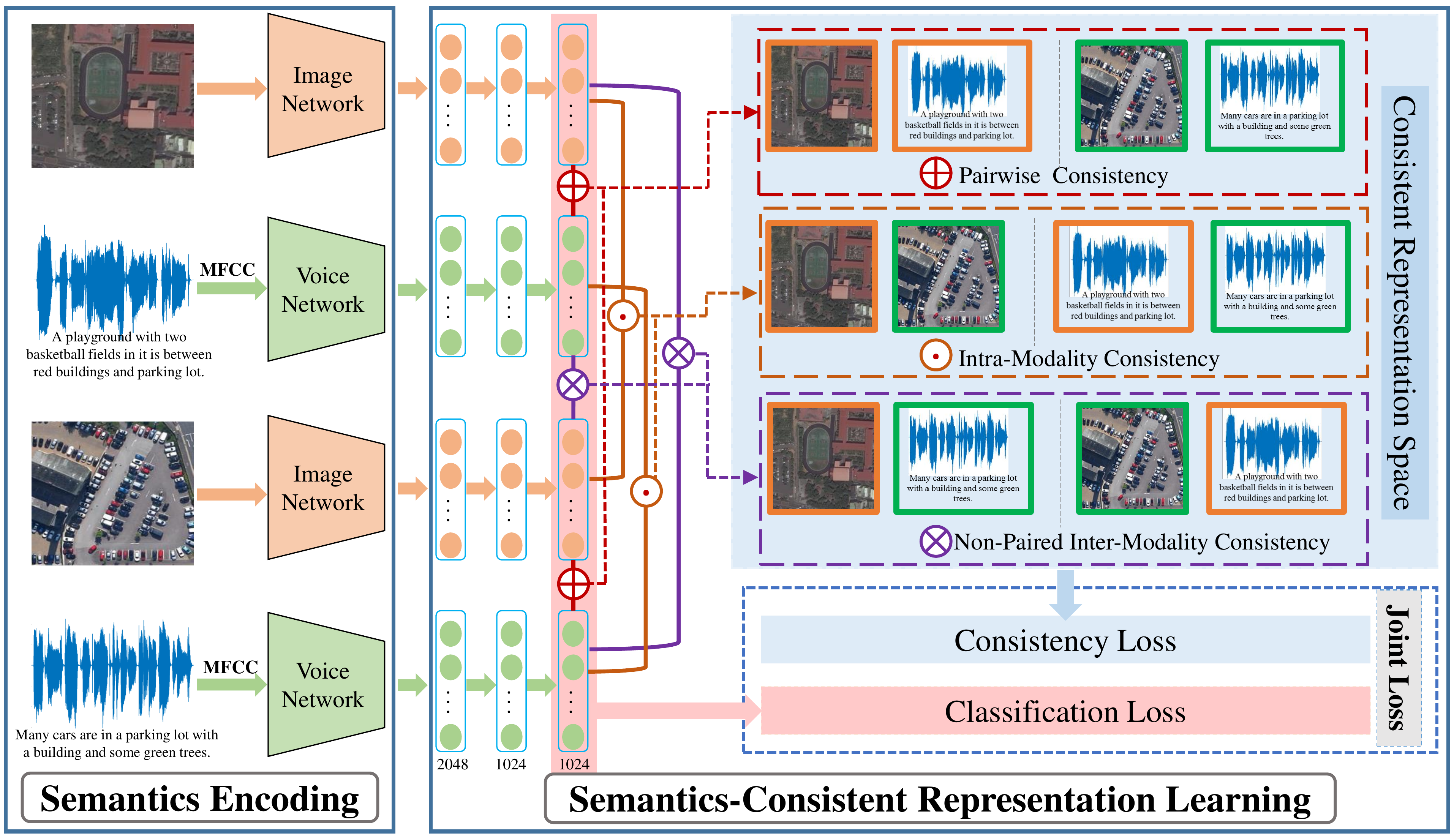}
\renewcommand{\figurename}{Fig.}
\end{center}
    \caption{\small{The overall framework of the proposed SCRL method. Firstly, the image encoding network is adopted to extract high-level image features, and the voice encoding network is devised to obtain high-level voice features. Secondly, the extracted image features and voice features are fed into the representation step to explore the consistent representation space. By exploring the space, the heterogeneous semantic gap is narrowed and the semantics-consistent representations across two modalities are learned by comprehensively considering the pairwise consistency ($\bigoplus$), intra-modality consistency ($\bigodot$), and non-paired inter-modality consistency ($\bigotimes$). MFCC represents the operation to extract the Mel-Frequency Cepstral Coefficients feature of the input voice.}}
\label{fig:2}
\end{figure*}

Inspired by the discussion above, a semantics-consistent representation learning (SCRL) method is proposed to make full use of these three relationships for the RS image-voice retrieval. As depicted in Fig. \ref{fig:2}, the proposed SCRL method can be divided into two steps: 1) semantics encoding and 2) semantics-consistent representation learning. Firstly, the VGG16 network \cite{Simonyan2014Very} is adopted to automatically extract high-level RS image features with a transfer learning strategy. Meanwhile, the input voices are processed as Mel-Frequency Cepstral Coefficients (MFCC) features \cite{8534807} and input into a 1-D dilated convolutional network to automatically extract high-level voice features. Secondly, to explicitly compute the similarity between representations from different modalities, a semantics-consistent representation space is explored by considering three kinds of relationships simultaneously. The relationships include pairwise, intra-modality, and non-paired inter-modality relationships. They are measured by the consistency loss to narrow the heterogeneous semantic gap across two modalities. In addition, to further learn significantly discriminative and more compact semantic representations, a classification loss is adopted for each branch. Afterwards, the consistency loss and the classification loss are combined jointly to learn ultimate semantics-consistent representations for the RS image-voice retrieval.

To sum up, the main contributions can be summarized into threefolds:

\begin{itemize}
  \item A novel RS image-voice retrieval method, SCRL, is proposed to fully explore the pairwise, intra-modality, and non-paired inter-modality relationships simultaneously. Then, semantics-consistent representations across the two modalities can be learned to effectively alleviate the pervasive heterogeneous semantic gap.

  \item An effective voice encoding network is proposed to learn high-level voice features. The network can capture the long range correlation of voice signal by introducing the 1-D dilated convolutional kernel.

  \item Experimental results on three challenging RS image-voice datasets show that the proposed SCRL method can significantly improve the performance of the RS image-voice retrieval. Especially, the mAP value can be improved by nearly 9\% compared with the state-of-the-arts.
\end{itemize}

The remainder of this paper is arranged as follows: Section \ref{relatedworks} introduces the related works about the task of RS image-voice retrieval. Section \ref{TheProposedmethod} elaborates the proposed SCRL method. Section \ref{experiments} shows and analyzes the experimental results. Finally, Section \ref{conclutions} presents the conclusion.

\section{Related Works}\label{relatedworks}
In this section, the related works about this paper are primarily restricted on uni-modal and cross-modal RS data retrieval methods.
\subsection{Uni-Modal RS Data Retrieval Methods}
Uni-modal RS data retrieval methods aim to search similar RS data in the same modality. Some early works are based on the hand-crafted feature \cite{luo2008indexing, yang2013geographic, aptoula2014remote}. For instance, Luo {\it et al.} \cite{luo2008indexing} present a method by comparing multiple-resolution wavelet features for satellite images retrieval. Yang and Newsam \cite{yang2013geographic} leverage local invariant features for RS image retrieval. Aptoula \cite{aptoula2014remote} proposes to apply global morphological texture descriptors for RS image retrieval. However, these methods are not suitable for large-scale RS image retrieval since they are based on low-level hand-crafted features. With the rapid development of artificial intelligent technology \cite{ning2020audio, zhao2019weather, yuan2018remote, zhao2019cam, yuan2019spatial}, a large number of uni-modal RS data retrieval methods based on deep learning have emerged. For instance, Tang {\it et al.} \cite{tang2017two-stage} develop a two-stage re-ranking method to improve the retrieval performance. Shao {\it et al.} \cite{8954885} design a fully convolutional network to solve the problem of poor retrieval performance due to multiple labels for single images. Ye {\it et al.} \cite{ye2018remote} utilize weighted distance as similarity criteria to learn determinative representations for the RS image retrieval. To address the RS image retrieval problem due to multiple land-cover classes, Chaudhuri {\it et al.} \cite{chaudhuri2017multilabel} introduce a semi-supervised graph-theoretic method with only a small number of training images characterized by multi-labels. Kang {\it et al.} \cite{kang2020graph} propose a graph relation network to preserve the complex semantic relations pervading RS scenes for multi-label RS image retrieval. In response to the scarcity of labeled images, Liu {\it et al.} \cite{liu2020similarity} present an unsupervised deep transfer learning method based on the similarity learning for the RS image retrieval. Ye {\it et al.} \cite{ye2019sar} propose to extract the domain-invariant feature for the RS image retrieval using an unsupervised domain adaptation model. To obtain the quicker response during retrieval, several techniques consider binary features \cite{demir2016hashing-based, 8067633, song2020unified}. Song {\it et al.} \cite{song2020unified} develop an image retrieval and compression method based on binary generative adversarial network. Demir and Bruzzone \cite{demir2016hashing-based} introduce two kernel-based nonlinear hashing methods for the RS image retrieval to reduce the amount of memory required. Li {\it et al.} \cite{8067633} introduce a RS image retrieval method, which integrates the feature learning network and hashing network into a unified framework.

\subsection{Cross-Modal RS Data Retrieval Methods}
The objective of cross-modal RS data retrieval methods is to search similar RS data across different modalities. Due to the urgent demand for RS data analysis, cross-modal RS data retrieval attracts increasing attention from researchers. Li {\it et al.} \cite{8385104} propose an end-to-end training method to learn source-invariant features for the cross-source RS image retrieval. Xiong {\it et al.} \cite{9044737} present a discriminative distillation network to eliminate the inconsistency across different modalities. To develop an effective retrieval system in harsh environments, Xiong {\it et al.} \cite{xiong2020deep} design another deep cross-modality hashing network for the Optical and SAR RS images retrieval. Recently, a more natural cross-modal RS data retrieval method has emerged, which is based on RS images and human voices. As the first attempt for the task of RS image-voice retrieval task, Guo {\it et al.} \cite{8486338} devise a deep visual-audio network to learn the correspondence between RS images and voices. Subsequently, they develop another cross-modal RS image-voice retrieval method to further improve the retrieval performance \cite{8896932}. Based on the works of Guo {\it et al.}, Chen and Lu \cite{chen2019a} present to leverage the triplet similarity of deep features to solve the insufficient utilization problem for the relative semantic similarity relationships. In addition, Chen {\it et al.} \cite{9044618} integrate the feature learning and hash code learning into a unified framework to achieve fast retrieval property. As another implementation, Chaudhuri {\it et al.} \cite{chaudhuri2020cmir-net} propose a deep neural network to learn a common embedding of RS images and spoken words rather than complete spoken sentences. Song {\it et al.} \cite{song2013inter} adopt an inter-media hashing (IMH) model for exploring the data correlations from heterogeneous data sources.

This paper is dedicated to conduct the cross-modal RS image-voice retrieval by fully considering the semantic relationships among representations across different modalities. The proposed method comprehensively improves the pairwise, intra-modality, and non-paired inter-modality consistency so as to narrow the heterogeneous semantic gap and learn semantics-consistent representations for RS image-voice retrieval.

\section{The Proposed method}\label{TheProposedmethod}
The proposed SCRL method is composed of two steps: 1) semantics encoding and 2) semantics-consistent representation learning. In this section, the overall framework of the proposed method is described firstly. Secondly, the semantics encoding and semantics-consistent representation learning steps are elaborated, respectively. Thirdly, the loss functions are presented. Finally, the optimizing strategy is introduced.

\subsection{Overall Framework}\label{Framework}
As shown in Fig. \ref{fig:2}, the proposed SCRL method consists of four parallel branches and can be divided into two steps, including semantics encoding and semantics-consistent representation learning steps. Two of the branches are related to the image modality, and the rest are related to the voice modality. The branches, which are responsible for the same modality, share the same parameters. As a consequence, the features from the same modality own the same transforms. After each iteration, four features from the two modalities are obtained by the semantics encoding step. Then, the obtained features are used as the input of the semantics-consistent representation learning step to learn four representations. During this step, the relationships of different representations are explored in a defined consistent representation space. The relationships include pairwise, intra-modality and non-paired inter-modality relationships. All the relationships are sought by calculating the distance between two representations to preserve the semantic consistency across different modalities. Specifically, the pairwise relationship is built to make the semantic information be consistent for the paired RS image-voice. The intra-modality relationship is considered to keep the semantic consistency within each modality. The non-paired inter-modality relationship is developed to enforce the semantic information to be consistent across different modalities for unpaired samples. Afterwards, the three kinds of relationships are measured by a consistency loss. Finally, we adopt a joint loss function to simultaneously minimize the consistency loss of the consistent representation space and the classification loss of each branch.

The formal definitions in this paper are given as follows. Let $\mathcal D = \left\{ {\left( {{\bf{X}}_m^I,{\bf{X}}{}_m^V,{{\bf{y}}_m}} \right)} \right\}_{m = 1}^N$ be the collection of $N$ instances of RS image-voice pairs, where ${\bf{X}}_m^I$ is the $m$-th input image sample, ${\bf{X}}_m^V$ is the $m$-th input voice sample, and ${{\bf{y}}_m}$ is the $m$-th semantic label. The goal of the semantics encoding step is to learn two mapping functions $\mathcal F_{I}$ and $\mathcal F_{V}$ for extracting high-level image features $\left\{{\bf{s}}_m^I\right\}_{m = 1}^N$ and voice features $\left\{{\bf{s}}_m^V\right\}_{m = 1}^N$. Afterwards, the semantics-consistent representation learning step aims to learn two mapping functions $\mathcal R_{I}$ and $\mathcal R_{V}$ for obtaining the semantics-consistent image representations $\left\{{\bf{\varphi}}_m^I\right\}_{m = 1}^N$ and voice representations $\left\{{\bf{\varphi}}_m^V\right\}_{m = 1}^N$. The obtained semantics-consistent representations are exploited for RS image-voice retrieval.

\subsection{Semantics Encoding}\label{FeatureExtraction}
As shown in Fig. \ref{fig:2}, the semantics encoding step consists of four parallel networks, in which the two image encoding networks share the same parameters and the two voice encoding networks share the same parameters. The details about image encoding networks and voice encoding networks are described as follows.

\subsubsection{\textbf{Image Semantics Encoding}}
According to the previous works \cite{Noh2017Large, lu2018hierarchical}, the high-quality feature is quite important for the retrieval task. CNN features have been proved to be high-efficiency for the retrieval task because of the success in expressing the high-level semantic information. To learn high-quality CNN features, massive amounts of data with manual annotations, such as ImageNet, are necessary. Unfortunately, large-scale dataset with manual annotation in RS domain is unavailable. In order to solve this problem, a transfer learning strategy has emerged. The transfer learning strategy applies knowledge learned from one domain to other domains \cite{2018Large}. In this paper, the transfer learning strategy is leveraged to extract high-quality image features. To be more specific, the VGG16 network \cite{Simonyan2014Very} pre-trained on ImageNet is adopted to acquire the RS image features since some useful texture features may share in natural images and RS images \cite{chen2018end-to-end}.

When using the pre-trained VGG16 network for image semantics encoding, the fully connected layers are removed to obtain high-level image features. Following the previous work \cite{gatys2016image}, the network is normalized by scaling weights to ensure the mean activation of each convolutional filter over images and positions to be one. Specifically, the mean activation $s_{i}^{l}$ of each convolutional filter in the $l$-th layer over the whole training image set $\mathcal X$ and all $M_{l}$ spatial locations can be calculated as follows:
\begin{equation}\label{Eq.:1}
s_{i}^{l}=\frac{1}{NM_{l}}\sum_{\mathcal X}\sum_{j=1}^{M_{l}}{\rm relu}(w_{i}^{l}\ast x_{j}^{l-1}+b_{i}^{l}),
\end{equation}
where $w_{i}^{l}$ represents the weight of $i$-th convolutional filter in the $l$-th layer, $x_{j}^{l-1}$ stands for the $j$-th patch in the $(l-1)$-th layer, $\ast$ indicates the convolutional operation, $b_{i}^{l}$ denotes the bias term in the $l$-th layer, and $N$ is the total number of the training images. Then, the parameters $w_{i}^{l}$ and $b_{i}^{l}$ are scaled by $\frac{1}{s_{i}^{l}}$ as follows:
\begin{equation}\label{Eq.:2}
\mathbb{E}_{{\mathcal X},j}[{\rm relu}(\frac{w_{l}^{i}}{s_{i}^{l}}\ast x_{j}^{l-1}+\frac{b_{l}^{i}}{s_{i}^{l}})]=1.
\end{equation}
According to the above process, the normalization is conducted from bottom layers to top layers sequentially.
The process of image semantics encoding can be written as:
\begin{equation}\label{Eq.:3}
{\bf{s}}_m^I=\mathcal F_{I}({\bf{X}}_m^I;\theta_{I}),
\end{equation}
where ${\bf{s}}_m^I \in \mathbb{R}^{h_1\times w_1\times c_1}$ represents the extracted semantics feature of the $m$-th image ${\bf{X}}_m^I$, $\mathcal F_{I}$ indicates the mapping function from image to the corresponding feature, and $\theta_{I}$ is the parameter.

\subsubsection{\textbf{Voice Semantics Encoding}} \label{VoiceFeatureExtraction}
Voice belongs to a continuous one-dimensional signal and the word in it cannot be distinguished easily by computer. In order to quantify the continuous voice, Mel-Frequency Cepstral Coefficients (MFCC) feature is first extracted to representant the voice signal according to previous works \cite{8534807, chowdhury2020fusing}. Here, the MFCC feature is expressed as ${\bf{X}}_m=\Upsilon({\bf{X}}_m^V)$.

The obtained MFCC feature is reshaped, and then fed into the voice encoding network to perform high-level voice features. To capture long range correlations, 1-D dilated convolutional kernels are adopted to construct the voice network. Specifically, the voice network is composed of 5 cascaded 1-D convolutional layers and pooling layers, where each convolutional layer adopts dilated convolutional kernel. The flattened MFCC feature is fed into the voice network and the learned high-level voice feature can be denoted as:
\begin{equation}\label{Eq.:4}
{\bf{s}}_m^V=\mathcal F_{V}({\bf{X}}_m;\theta_{V}),
\end{equation}
where ${\bf{s}}_m^V \in \mathbb{R}^{h_2\times w_2\times c_2}$ represents the extracted high-level feature of the $m$-th MFCC feature ${\bf{X}}_m$, $\mathcal F_{V}$ indicates the mapping function from MFCC feature to the corresponding high-level voice feature, and $\theta_{V}$ is the parameter.

\subsection{Semantics-Consistent Representation Learning}
This step aims to learn semantics-consistent representations by comprehensively considering the pairwise, intra-modality, and non-paired inter-modality relationships across different modalities. As shown in Fig. \ref{fig:2}, these three relationships are measured by a consistency loss to model the similarity of the learned representations. In addition, a classification loss is adopted for each branch to enhance the semantic discrimination ability of representations, resulting in more compact representations. The consistency loss and classification loss are combined jointly to learn semantics-consistent representations for the RS image-voice retrieval. Hereinafter, the details about the semantics-consistent representation learning are dwelled on.

\subsubsection{\textbf{The Architecture of Semantics-Consistent Representation Learning}}
As shown in Fig. \ref{fig:2}, the semantics-consistent representation learning step contains 4 parallel branches. Similar to the semantics encoding step, two of the branches are related to the image modality, and the rest are related to the voice modality. Actually, the branches responsible for the same modality shares the same weights. Each branch is constructed by three cascaded fully connected layers. The specific parameters of each branch are shown in Fig. \ref{fig:2}. The features from the semantics encoding step are taken as inputs for the semantics-consistent representation learning step. Then, the representations are output from each branch in the representation learning step.

As for the image branch, the learned image representation can be expressed as:
\begin{equation}\label{Eq.:5}
{\bf{\varphi}}_m^I=\mathcal R_{I}({\bf{s}}_m^I;\Theta_{I}),
\end{equation}
where ${\bf{\varphi}}_m^I \in \mathbb{R}^{d_1}$ represents the learned semantics-consistent representation for the $m$-th image, $\mathcal R_{I}$ indicates the mapping function from the image feature to the corresponding semantics-consistent representation, and $\Theta_{I}$ is the parameter.

As for the voice branch, the learned voice representation can be expressed as:
\begin{equation}\label{Eq.:6}
{\bf{\varphi}}_m^V=\mathcal R_{V}({\bf{s}}_m^V;\Theta_{V}),
\end{equation}
where ${\bf{\varphi}}_m^V \in \mathbb{R}^{d_2}$ represents the learned semantics-consistent representation for the $m$-th voice, $\mathcal R_{V}$ indicates the mapping function from the voice feature to the corresponding semantics-consistent representation, and $\Theta_{V}$ is the parameter.

\subsubsection{\textbf{The Pairwise Consistency}}
The pairwise consistency represents the semantic information of representations from the paired RS image-voice should be consistent since they describe the same semantic concept in the applied RS image-voice datasets. To measure the similarity of representations, the cosine distance is adopted as it is commonly used in cross-modal retrieval \cite{Chen2020Deep, 8985543}. The cosine distance can be defined as:
\begin{equation}\label{Eq.:7}
D(\bf{x}, \bf{y})=1 - \frac{{\bf{x} \cdot \bf{y}}}{{\left\| \bf{x} \right\|\left\| \bf{y} \right\|}},
\end{equation}
where $\bf{x}$ and $\bf{y}$ represent two vectors with the same length, and $\cdot$ indicates the operation of dot product between the two vectors.

Based on the cosine distance, the pairwise consistency can be maintained by the pairwise consistency loss, which is defined as:
\begin{equation}\label{Eq.:8}
\mathcal{L}_{pair} = D(\varphi_{i}^{I}, \varphi_{i}^{V}),
\end{equation}
where $\varphi_{i}^{I}$ and $\varphi_{i}^{V}$ denote representations of the $i$-th image and voice, respectively. Minimizing the pairwise consistency loss leads to a common space, where representations describing the same semantic concept are clustered together and the pairwise consistency is preserved.

\subsubsection{\textbf{The Intra-Modality Consistency}}
The intra-modality consistency measures the relationships between two representations from the same modality and belonging to the same semantic concept. In order to preserve the intra-modality consistency, an intra-modality consistency loss is introduced, which is defined as:
\begin{equation}\label{Eq.:9}
\begin{split}
\mathcal{L}_{intra}& = \hbar\left(1-\ell_{ij}\left(\xi-D\left(\varphi_{i}^{I}, \varphi_{j}^{I}\right)\right)\right)\\
&+ \hbar\left(1-\ell_{ij}\left(\xi-D\left(\varphi_{i}^{V}, \varphi_{j}^{V}\right)\right)\right),
\end{split}
\end{equation}
where $\hbar(x)=max(0,x)$ is the hinge function, $\xi$ is the pre-defined margin threshold. $\ell_{ij}$ denotes the label indicator. If $\varphi_{i}^{I}$ and $\varphi_{j}^{I}$, or $\varphi_{i}^{V}$ and $\varphi_{j}^{V}$ describe the same semantic concept, $\ell_{ij}=+1$; otherwise, $\ell_{ij}=-1$. It is to noted that $i\neq j$. Minimizing the intra-modality consistency loss results in a common space, where representations describing the same semantic concept within the same modality are enforced to be close, while representations describing different semantic concepts within the same modality are enforced to be far.

\subsubsection{\textbf{The Non-Paired Inter-Modality Consistency}}
The non-paired inter-modality consistency measures the relationships between two representations from different modalities but belonging to the same semantic concept. Since the pairwise loss only model the relationships of paired image-voice pairs with the same semantic concept in the applied RS image-voice datasets, the relationships between other non-paired data cannot been explored effectively, and the compact representations for different sematic concepts are unable to be learned. To this end, the inter-modality consistency loss is used to make up for the lack of the pairwise loss. To maintain the non-paired inter-modality consistency, the inter-modality consistency loss is defined as:

\begin{equation}\label{Eq.:10}
\begin{split}
\mathcal{L}_{inter} &= \hbar\left(1-\ell_{ij}\left(\zeta-D\left(\varphi_{i}^{V}, \varphi_{j}^{I}\right)\right)\right)\\
&+ \hbar\left(1-\ell_{ij}\left(\zeta-D\left(\varphi_{i}^{I}, \varphi_{j}^{V}\right)\right)\right),
\end{split}
\end{equation}
where $\zeta$ is the pre-defined margin threshold. If $\varphi_{i}^{V}$ and $\varphi_{j}^{I}$, or $\varphi_{i}^{I}$ and $\varphi_{j}^{V}$ describe the same semantic concept, $\ell_{ij}=+1$; otherwise, $\ell_{ij}=-1$. Minimizing the inter-modality consistency loss leads to a common space, where representations describing the same semantic concept across the two modalities are enforced to be close, while representations describing different semantic concepts across the two modalities are enforced to be far.

\subsection{Loss Function}
\subsubsection{\textbf{The Consistency Loss}}
The consistency loss is defined in the consistent representation space to model the similarity of the learned representations, which combines the pairwise, intra-modality and non-paired inter-modality consistency mentioned above. As a consequence, the consistent representation space possesses all the advantages which are contained in all three common spaces obtained by pairwise, intra-modality and inter-modality consistency loss, respectively. Considering that the intra-modality and inter-modality loss are calculated from the same two pairs of samples, the significance of the intra-modality loss and the inter-modality consistency loss is set equal. Therefore, the consistency loss in the consistent representation space is defined as:
\begin{equation}\label{Eq.:11}
\mathcal{L}_{consi} = \mathcal{L}_{pair}+\eta_1(\mathcal{L}_{intra}+\mathcal{L}_{inter}),
\end{equation}
where $\eta_1$ is the trade-off coefficient controlling the contribution of the last two terms.

\subsubsection{\textbf{The Classification Loss}}
As for the representation on the top of each branch, a classification process is added to mine the intrinsic semantic information in each image and voice and further to model the discrimination and compactness of the learned representations. Specifically, a softmax layer is adopted for each image branch as:
\begin{equation}\label{Eq.:12}
p_m^I = {\rm softmax}({\bf{W}}^I\varphi_m^I+{\bf{b}}^I),
\end{equation}
where $p_m^I$ is the probability of belonging to the corresponding semantic concept for the $m$-th image. ${\rm softmax}$ represents the softmax activation function. ${\bf{W}}^I$ and ${\bf{b}}^I$ are the parameters in the softmax layer. Similarly, a softmax layer is adopted for each voice branch as:
\begin{equation}\label{Eq.:13}
p_m^V = {\rm softmax}({\bf{W}}^V\varphi_m^V+{\bf{b}}^V),
\end{equation}
where $p_m^V$ is the probability of belonging to the corresponding semantic concept for the $m$-th voice. ${\bf{W}}^V$ and ${\bf{b}}^V$ are the parameters in the softmax layer.

Then, the classification loss can be defined as:
\begin{equation}\label{Eq.:14}
\begin{split}
\mathcal{L}_{class} = - y_{it}{\rm log}(p_{it}^I+\epsilon)-y_{it}{\rm log}(p_{it}^V+\epsilon),
\end{split}
\end{equation}
where $y_{it}$ is the true semantic label of the $i$-th sample, where $t$ indexes the $t$-th class. $p_{it}^I$ is the predicted probability distribution of the $i$-th images. $p_{it}^V$ is the predicted probability distribution of the $i$-th voices, where $t$ indexes the $t$-th class. $\epsilon$ represents a regularization constant to avoid the NaN value in the loss. By minimizing the classification loss function, the semantic discrimination ability of the common representations in the consistent representation space can be greatly enhanced.

\begin{algorithm}[htb]
\renewcommand{\algorithmicrequire}{\textbf{Input:}}
\renewcommand\algorithmicensure {\textbf{Output:} }

\caption{The proposed SCRL method}
\label{alg:Framwork}
\begin{algorithmic}[1]

\REQUIRE ~~\\
Training collection $\mathcal D = \left\{ {\left( {{\bf{X}}_m^I,{\bf{X}}{}_m^V,{{\bf{y}}_m}} \right)} \right\}_{m = 1}^N$ of RS image-voice pairs;\\
Learning rate $lr$, trade-off coefficients $\eta_1$ and $\eta_2$, iterative epoch $K$.

\ENSURE ~~\\
The parameters $\theta_I$ and $\theta_V$ in the feature extraction step;\\
The parameters $\Theta_I$ and $\Theta_V$ in the representation learning step.

\renewcommand{\algorithmicrequire}{\textbf{Initialization:}}
\REQUIRE ~~\\
The parameter $\theta_I$ is initialized by the pre-trained VGG16 network;\\
The parameters $\theta_V$, $\Theta_I$ and $\Theta_V$ are randomly initialized by truncated\_normal distribution.

\renewcommand{\algorithmicrequire}{\textbf{Repeat:}}
\REQUIRE ~~\\
\STATE Sample the input RS image-voice pairs $\left({{{\bf{X}}_i^I,{\bf{X}}{}_i^V,{{\bf{y}}_i}}}\right)$ and $\left({{{\bf{X}}_j^I,{\bf{X}}{}_j^V,{{\bf{y}}_j}}}\right)$;

\STATE Calculate the image features ${\bf{s}}_i^{I}$ and ${\bf{s}}_j^{I}$ via Eq. \ref{Eq.:3}, and voice features ${\bf{s}}_i^{V}$ and ${\bf{s}}_j^{V}$ according to Eq. \ref{Eq.:4};

\STATE Learn the semantics-consistent representations $\varphi_{i}^{I}$ and $\varphi_{j}^{I}$ according to Eq. \ref{Eq.:5}, as well as $\varphi_{i}^{V}$ and $\varphi_{j}^{V}$ via Eq. \ref{Eq.:6};

\STATE Calculate the joint loss $\mathcal{L}_{joint}$ according to Eq. \ref{Eq.:16};

\STATE Update the parameters $\theta_V$, $\Theta_I$ and $\Theta_V$ by utilizing RMSProp.

\renewcommand\algorithmicensure {\textbf{Until:}{\,}{A fixed iterative epoch $K$ or convergence.}}
\ENSURE ~~\\
\renewcommand\algorithmicensure {\textbf{Return:}{\,}{$\theta_I$, $\theta_V$, $\Theta_I$ and $\Theta_V$}}
\ENSURE ~~\\
\end{algorithmic}

\end{algorithm}

\subsubsection{\textbf{The Joint Loss}}
With the above definition, a joint loss function is raised to simultaneously calculate the consistency loss of the consistent representation space and the classification loss of each branch. The joint loss function can be written as:
\begin{equation}\label{Eq.:15}
\mathcal{L}_{joint} = \mathcal{L}_{consi}+\eta_2\mathcal{L}_{class},
\end{equation}
where $\eta_2$ is the trade-off coefficient controlling the contribution of the second term. By combining Eq. \ref{Eq.:11} and Eq. \ref{Eq.:15}, the joint loss can be rewritten as:
\begin{equation}\label{Eq.:16}
\mathcal{L}_{joint} = \mathcal{L}_{pair}+\eta_1(\mathcal{L}_{intra}+\mathcal{L}_{inter})+\eta_2\mathcal{L}_{class}.
\end{equation}
By minimizing the joint loss, the semantics-consistent representations are learned for the RS image-voice retrieval.

\subsection{Optimizing Strategy}
Based on the joint loss, the proposed SCRL method is optimized as follows. The parameter $\theta_I$ in the image encoding network is initialized by VGG16 network pre-trained on the ImageNet \cite{chen2020deep_PAIR}. The parameters $\theta_V$ in the voice encoding network, $\Theta_I$ and $\Theta_V$ in the semantics-consistent representation step are randomly initialized by truncated\_normal distribution \cite{yuan2019bio-inspired}. In each training iteration, the optimizing process can be divided into five main steps. Firstly, we sample two pairs of RS image-voice from the training collection of RS image-voice pairs. Secondly, the sampled RS image-voice pairs are input into the semantics encoding step to obtain image and voice features according to Eq. \ref{Eq.:3} and Eq. \ref{Eq.:4}, respectively. Thirdly, the extracted image and voice features are taken as input of the semantics-consistent representation step for learning semantics-consistent representations via Eq. \ref{Eq.:5} and Eq. \ref{Eq.:6}. Fourthly, the joint loss is calculated according to Eq. \ref{Eq.:16}. Finally, the parameters $\theta_V$, $\Theta_I$ and $\Theta_V$ are updated by minimizing the joint loss with the RMSprop optimization algorithm \cite{yuan20193g}. When a fixed iterative epoch, which is set as 50, is reached or the model is convergent, the optimizing process is terminated. Afterwards, the parameters $\theta_I$, $\theta_V$, $\Theta_I$ and $\Theta_V$ are leveraged to compute ultima semantics-consistent representations for the RS image-voice retrieval. Noted that the proposed SCRL method is trained end-to-end. The details about the optimization process of the proposed SCRL method are shown in Algorithm \ref{alg:Framwork}.

\section{Experiment and Results}\label{experiments}
In this section, experimental datasets, implementation details, evaluation metrics and parameter analysis are introduced. In addition, the experimental results are compared with some state-of-the-art methods and the ablation analysis is presented to prove the effectiveness of the proposed SCRL method.
\subsection{Datasets}
In order to verify the proposed SCRL method, three challenging RS image-voice datasets are applied, including Sydney, UCM and RSICD image-voice datasets \cite{8486338}. A brief introduction about the applied RS image-voice datasets is given as follows.

\subsubsection{\textbf{Sydney Image-Voice Dataset \cite{8486338}}} The Sydney image-voice (Sydney IV) dataset contains 613 RS images and 3065 voices of 7 classes, where each image corresponds to five different voices. In this paper, we follow the previous works \cite{8896932, chen2019a, 9044618}, and randomly sample a voice from the five voices for each image to construct RS image-voice pairs. As for the data partitioning, 80\% RS image-voice pairs are randomly selected for training, and the rest 20\% are selected for testing.

\subsubsection{\textbf{UCM Image-Voice Dataset \cite{8486338}}} The UCM image-voice (UCM IV) dataset includes 2100 RS images and 10500 voices, where each image corresponds to five different voices. Note that the dataset can be divided into 21 classes, where each class includes 100 images and 500 voices. In this paper, we follow the previous works \cite{8896932, chen2019a, 9044618}, and randomly sample a voice from the five voices for each image to construct RS image-voice pairs. As for the data partitioning, 80\% RS image-voice pairs are randomly selected as training set, and the rest 20\% are selected as testing set.

\subsubsection{\textbf{RSICD Image-Voice Dataset \cite{8486338}}} The RSICD image-voice (RSICD IV) dataset involves 10921 RS images and 54605 voices of 30 classes, where each image corresponds to five different voices. In this paper, we follow the previous works \cite{8896932, chen2019a, 9044618}, and randomly sample a voice from the five voices for each image to construct RS image-voice pairs. As for the data partitioning, 80\% RS image-voice pairs are randomly selected for training, and the rest 20\% are selected for testing.

\subsection{Implementation Details}
In this work, the proposed SCRL method contains four branches, in which two branches are responsible for the image modality, and the others are responsible for the voice modality. As for branches, responsible for the same modality, the parameters are sharing. The specific architecture of the voice encoding network is reported in TABLE \ref{tab1}. The dilation rate in the first convolutional layer is set as 3 and the others are set as 2. The input images are uniformly adjusted to 224$\times$224. Before input into the voice network, the voices are resampled at 22050Hz and preprocessed as MFCC features by a window with size of 16 millisecond and shift of 5 millisecond. From the output of the semantics encoding step, the image features are with size of 7$\times$7$\times$512 (namely $h_1=w_1=7$ and $c_1=512$), and the voice features are with size of 375$\times$1$\times$48 (namely $h_2=375$, $w_2=1$ and $c_2=48$). Before input into the semantics-consistent representation learning step, the image features are adjusted as size of 512 by the operation of global average pooling \cite{7532802}, and the voice features are adjusted as size of 18000 by the operation of flattening. In the semantics-consistent representation step, $ReLU$ and $Tanh$ are applied as activation functions of image branches and voice branches, respectively \cite{Hu2019Deep, Shao2019Cloud}. The ultima semantics-consistent representations are with size of 1024 (namely $d_1=d_2=1024$).

The proposed method is optimized utilizing a RMSProp optimizer \cite{yuan2019bio-inspired}, in which the weight decay is set as 0.0005 and momentum is set to 0.9. The learning rate is set as 0.0004. The batch size is set to 32. The trade-off coefficients $\eta_1$ and $\eta_2$ are set as 1 and 0.1, respectively, which are determined by a grid search strategy \cite{9044618}. The experiment is conducted on the PC with a TITAN X (Pascal) GPU and 64G RAM.

\begin{table}
\caption{THE DETAILED ARCHITECTURE OF THE VOICE NETWORK.}
\begin{center}
\small
\setlength{\tabcolsep}{4.5mm}{
\begin{tabular}{cccc} \toprule
Layer    &Output-Size                 &Kernel                  &Stride              \\ \hline
Conv1    &24000$\times$1$\times$1     &7$\times$1$\times$1     &1$\times$1$\times$1 \\
Conv2    &12000$\times$1$\times$6     &3$\times$1$\times$6     &1$\times$2$\times$1 \\
Conv3    &6000$\times$1$\times$12     &3$\times$1$\times$12    &1$\times$2$\times$1 \\
Pool1    &3000$\times$1$\times$12     &1$\times$2$\times$1     &1$\times$2$\times$1 \\
Conv4    &1500$\times$1$\times$24     &3$\times$1$\times$24    &1$\times$2$\times$1 \\
Conv5    &750$\times$1$\times$48      &3$\times$1$\times$48    &1$\times$2$\times$1 \\
Pool2    &375$\times$1$\times$48      &1$\times$2$\times$1     &1$\times$2$\times$1 \\
\bottomrule
\end{tabular}}
\end{center}
 \label{tab1}%
\end{table}

\subsection{Evaluation Metrics and Comparison Methods}
In this work, two protocols are adopted, including using images as queried samples to retrieve the corresponding voices (I$\rightarrow$V) and using voices as queried samples to retrieve the corresponding images (V$\rightarrow$I). To evaluate the performance of the proposed SCRL method, two commonly used evaluation metrics are adopted, including mean average precision (mAP) \cite{khurshid2020a} and the precision of the top-k ranking result ($\bf P$@{\textbf k}) \cite{li2017partial}. The mAP score measures the mean value of average precision, which considers the precision and the returned ranking results at the same time. Precision is the ratio of the returened relevant samples to the queried samples. The $\bf P$@{\textbf k} score measures the precision of the top-k retrieved samples. In this paper, the $\bf P$@{\textbf k} score is reported when $k$ equals to 1, 5 and 10, which is denoted as $\bf P$@{\textbf 1}, $\bf P$@{\textbf 5} and $\bf P$@{\textbf 10}, respectively. In addition, the precision curve is shown when the number of retrieved samples changes to further evaluate the proposed SCRL method.

In order to assess the effectiveness of the proposed SCRL method, 7 comparison methods are adopted, including SIFT+M \cite{Zhao2018SIFT}, DBLP \cite{DBLP:conf/nips/HarwathTG16}, CNN+SPEC \cite{8237335}, DVAN \cite{8486338}, CMIR-NET \cite{chaudhuri2020cmir-net}, DIVR \cite{9044618}, and DTBH \cite{chen2019a} methods. The SIFT+M method \cite{Zhao2018SIFT} leverages SIFT features of images and MFCC features of voices to perform the RS image-voice retrieval. The DBLP method \cite{DBLP:conf/nips/HarwathTG16} adopts an unsupervised manner to learn the coherence between audio and visual modalities. The CNN+SPEC method \cite{8237335} seamlessly unifies the learning of different modalities in an unsupervised manner. The DVAN method \cite{8486338} presents a novel image-voice learning framework to learn the cross-modal similarity. The CMIR-NET method \cite{chaudhuri2020cmir-net} proposes to learn the discriminative shared feature space of the input data for RS image-voice retrieval. The DIVR method \cite{9044618} leverages multi-scale context information guiding the low-dimensional hash code for RS image-voice retrieval. The DTBH method \cite{chen2019a} establishes a deep triplet-based hashing method, which integrates the hash code learning and the representation learning into a unified framework. These methods are implemented in this work.
It is to note that the hash-based retrieval methods, including DTBH and DIVR methods, use a 64-bit hash code for comparison. The experimental results and the corresponding analysis are given as follows.

\begin{figure}[tp]
\begin{center}
\includegraphics[width=\linewidth]{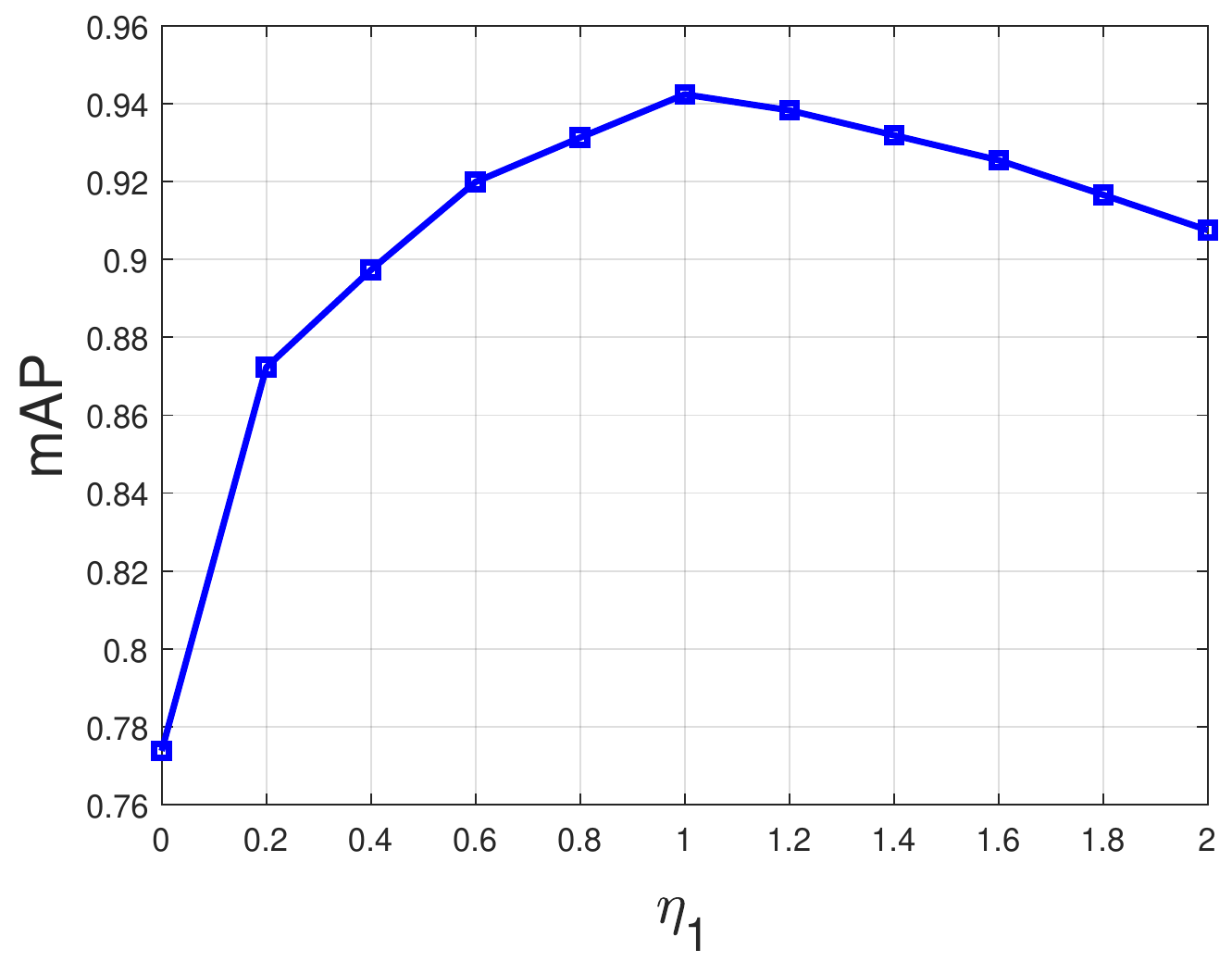}
\renewcommand{\figurename}{Fig.}
\vspace{-8 mm}
\end{center}
    \caption{\small{The retrieval performance with different values for $\eta_1$ by the I$\rightarrow$V protocol on the Sydney IV dataset.}}
\label{fig:3}
\end{figure}

\begin{figure}[tp]
\begin{center}
\includegraphics[width=\linewidth]{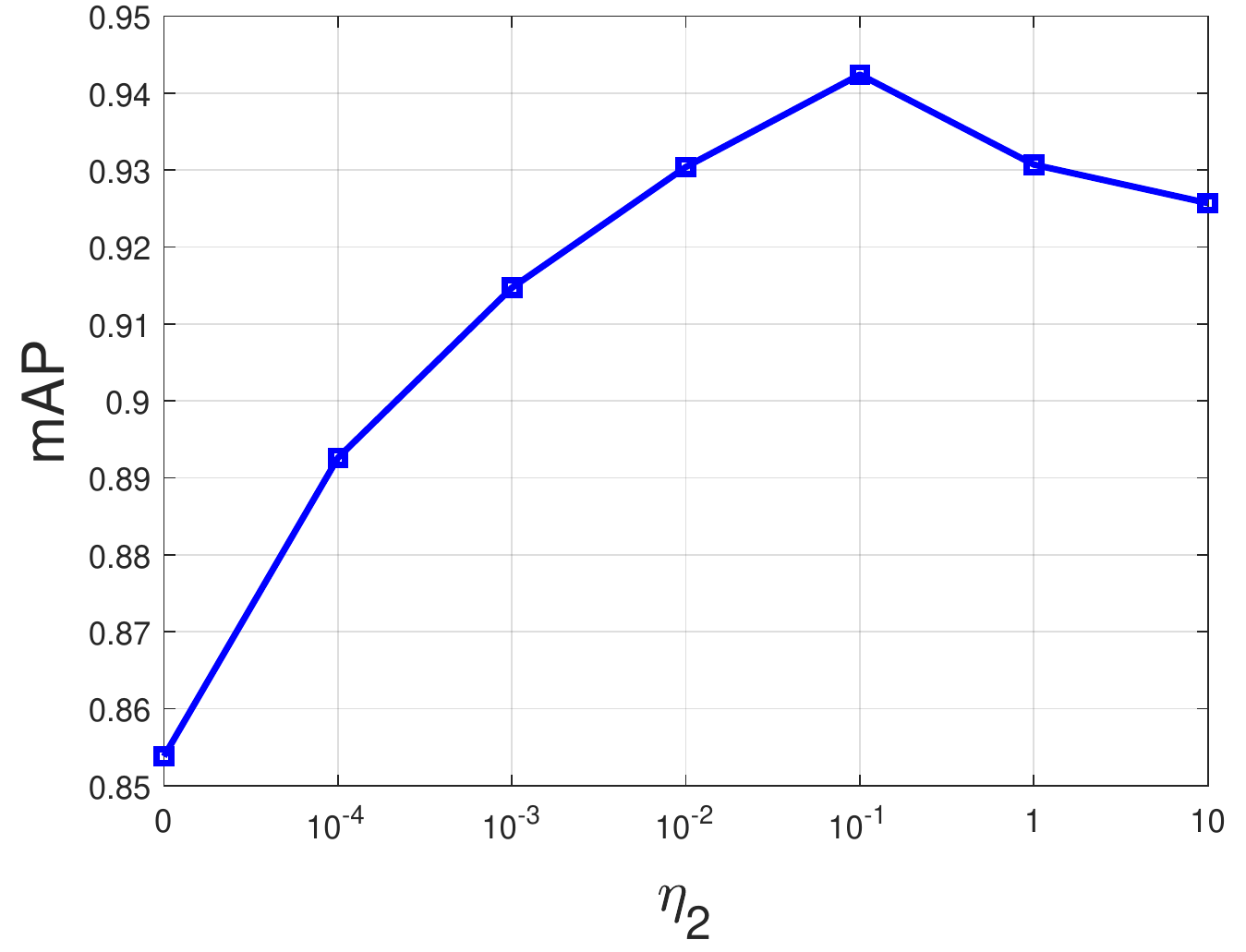}
\renewcommand{\figurename}{Fig.}
\vspace{-8 mm}
\end{center}
    \caption{\small{The retrieval performance with different values for $\eta_2$ by the I$\rightarrow$V protocol on the Sydney IV dataset.}}
\label{fig:4}
\end{figure}

\subsection{Parameter Analysis}
The hyperparameter $\eta_1$ controls the contributions of the intra-modality and non-paired inter-modality consistency, and $\eta_2$ controls the contribution of modeling the semantic classification. To study the performance impacts of them, we conduct parameter experiments about the two parameters with the I$\rightarrow$V protocol on the Sydney IV dataset. To this end, a grid search strategy is adopted to tune the two parameters following the previous work \cite{wang2017adversarial}.  Specially, the range of $\eta_1$ and $\eta_2$ are set as $\{0, 0.2, 0.4, 0.8, 1.0, 1.2, 1.4, 1.6, 1.8, 2.0\}$ and $\{0, 10^{-4}, 10^{-3}, 10^{-2}, 10^{-1}, 10\}$, respectively. Note that $\eta_1=0$ represents the SCRL with $\mathcal{L}_{pair}$ and $\mathcal{L}_{class}$ only, and $\eta_2=0$ represents the SCRL with $\mathcal{L}_{consi}$ only. The parameter experiments are conducted by changing one hyperparameter ({\it e.g.}, $\eta_1$) while fixing the other ({\it e.g.}, $\eta_2$). The detailed implementation and analysis are as follows.

Firstly, the value of $\eta_2$ is fixed at 1 empirically, $\eta_1$ is changed from 0 to 2 with increment 0.2 per step. The corresponding mAP scores about the parameter $\eta_1$ are recorded and shown in Fig. \ref{fig:3}. From the figure, it can be seen that both lower and higher values for $\eta_1$ result in lower mAP value, and the best mAP is obtained when $\eta_1$ equals to 1. Notice that when $\eta_1=0$, the retrieval performance of the proposed method is significantly reduced, because the neglect of the intra-modality consistency and non-paired inter-modality consistency leads to the same semantic concept across the two modalities not being matched correctly.

Secondly, the value of $\eta_1$ is fixed at 1, $\eta_2$ is changed from 0 to 10 with a ten times increment each step. Fig. \ref{fig:4} shows the performance for different $\eta_2$ values. From the figure, we can observe the best performance is obtained when $\eta_2=0.1$. It is noted that when $\eta_2$ equals to 0, the retrieval performance gets worse slightly, which indicates the classification loss contributes to enhancing the semantic discrimination ability and compactness of the semantics-consistent representations.

As a result, the trade-off coefficients $\eta_1$ and $\eta_2$ in the proposed SCRL method are set to 1 and 0.1, respectively.

\subsection{Ablation Analysis}
In this subsection, five variations of the proposed SCRL method are conducted on three challenging RS image-voice datasets to examine: 1) the importance of the pairwise consistency loss; 2) the impact of the intra-modality consistency loss; 3) the importance of the inter-modality consistency loss; 4) the importance of the classification loss; 5) the effect of the dilated convolutional kernel. The details about the implementation of different variations are elaborated as follows. Firstly, the variation of the proposed SCRL method without the pairwise consistency loss (SCRL-Pair) is implemented to verify the effect of the pairwise consistency loss. Secondly, the intra-modality consistency loss is discarded evolving into a new variation (SCRL-Intra). Thirdly, we omit the inter-modality consistency loss obtaining another new variation (SCRL-Inter) to determine the importance of the non-paired inter-modality relationship. Fourthly, the classification loss is abandoned evolving into the variation SCRL-Class. Finally, the dilation rate in the voice network is set as 1 generating the new variation SCRL-Dilation. The results of these variations are reported in TABLE \ref{tab:2}-\ref{tab:4}. According to the results, we give the analysis as the following aspects.

\renewcommand\arraystretch{1.1}
\begin{table}
  \centering
\small
  \caption{THE COMPARISON RESULTS BETWEEN THE PROPOSED SCRL METHOD AND OTHER METHIDS ON THE SYDNEY IV DATASET.}
  \begin{threeparttable}
 \setlength{\tabcolsep}{1.5mm}{	
    \begin{tabular}{ccccccc}\toprule
     Protocols                        &Methods         & mAP            & $\bf P$@{\textbf 1} &$\bf P$@{\textbf 5} &$\bf P$@{\textbf 10} \\
    \hline
    \multirow{13}*{I$\rightarrow$V}      &SIFT+M \cite{Zhao2018SIFT}               &31.67     &11.21     &35.00     &37.59      \\
                                         &DBLP \cite{DBLP:conf/nips/HarwathTG16}   &44.38     &56.51     &52.65     &49.68      \\
                                         &CNN+SPEC \cite{8237335}                  &46.67     &58.62     &55.00     &51.64      \\
                                         &DVAN \cite{8486338}                      &71.77     &75.86     &73.62     &72.93      \\
                                         &CMIR-NET \cite{chaudhuri2020cmir-net}    &78.44     &84.68     &82.54     &81.04      \\
                                         &DIVR \cite{9044618}                      &81.35     &88.26     &86.35     &84.47      \\
                                         &DTBH \cite{chen2019a}                    &92.45     &97.41     &95.63     &93.78      \\
         \cline{2-6}
                                         &SCRL-Pair                                &75.80      &80.11     &75.07     &71.90      \\
                                         &SCRL-Intra                               &79.79     &85.43     &81.20     &74.78      \\
                                         &SCRL-Inter                               &80.19     &86.24     &82.50     &76.57\\
                                         &SCRL-Class                               &85.38     &90.87     &88.11     &85.99\\
                                         &SCRL-Dilation                            &92.08     &93.68     &93.03     &91.65\\
                                         &\bf{SCRL}      &\textbf{94.24}    &\textbf{95.50}     & \textbf{95.17}  &\textbf{93.95}\\
    \hline
    \multirow{13}*{V$\rightarrow$I}      &SIFT+M \cite{Zhao2018SIFT}               &26.50     &34.48     &24.48     &23.28\\
                                         &DBLP \cite{DBLP:conf/nips/HarwathTG16}   &34.87     &21.63     &26.78     &30.94\\
                                         &CNN+SPEC \cite{8237335}                  &35.72     &17.24     &27.76     &31.21\\
                                         &DVAN \cite{8486338}                      &63.88     &67.24     &63.34     &67.07\\
                                         &CMIR-NET \cite{chaudhuri2020cmir-net}    &71.28     &76.69     &74.52     &71.60\\
                                         &DIVR \cite{9044618}                      &75.97     &80.44     &78.05     &76.27\\
                                         &DTBH \cite{chen2019a}                    &87.49     &92.18     &90.36     &88.82\\

       \cline{2-6}
                                         &SCRL-Pair                                &75.19     &78.67     &76.75     &75.12\\
                                         &SCRL-Intra                               &79.96     &85.37     &84.55     &83.17\\
                                         &SCRL-Inter                               &80.52     &85.99     &85.15     &83.69\\
                                         &SCRL-Class                               &84.87     &88.62     &87.64     &86.75\\
                                         &SCRL-Dilation                            &91.14     &92.68     &89.59     &89.11\\
                                         &\bf{SCRL}      &\textbf{93.53}    &\textbf{94.31}      &\textbf{91.38}  &\textbf{90.00}\\
\bottomrule
    \end{tabular}}%
  \label{tab:2}%
\end{threeparttable}
\end{table}%

\subsubsection{\textbf{Pairwise Consistency}} The pairwise consistency loss aims to narrow the distance between the representations from the RS image-voice pair. From the results in TABLE \ref{tab:2}-\ref{tab:4}, we can intuitively find that the retrieval performance is greatly improved when the pairwise consistency is taken into account. Concretely, compared with the method without the pairwise consistency loss (SCRL-Pair), the proposed SCRL method can improve the mAP value from 75.80\% to 94.24\%, from 35.25\% to 67.97\%, and from 15.80\% to 28.84\% for the I$\rightarrow$V protocol on Sydney, UCM and RSICD IV datasets, respectively. For the V$\rightarrow$I protocol, the mAP value is also improved greatly by the proposed SCRL method compared with the SCRL-Pair method. The improvements demonstrate that the pairwise consistency loss is effective for learning semantics-consistent representations.
\subsubsection{\textbf{Intra-Modality Consistency}} The intra-modality consistency loss aims to narrow the distance between two representations in the same modality to keep the semantic consistency of the same semantic concept. From the results in TABLE \ref{tab:2}-\ref{tab:4}, we can intuitively find that the retrieval performance is greatly improved when the intra-modality consistency is taken into account. The specific comparison situation corresponds to the results of SCRL-Intra method and the proposed SCRL method in TABLE \ref{tab:2}-\ref{tab:4}. The improvement is because the intra-modality consistency loss can constrain the model to shorten the distance between two representations from the same semantic concept in each modality.

\begin{figure}[tp]
\begin{center}
\includegraphics[width=\linewidth]{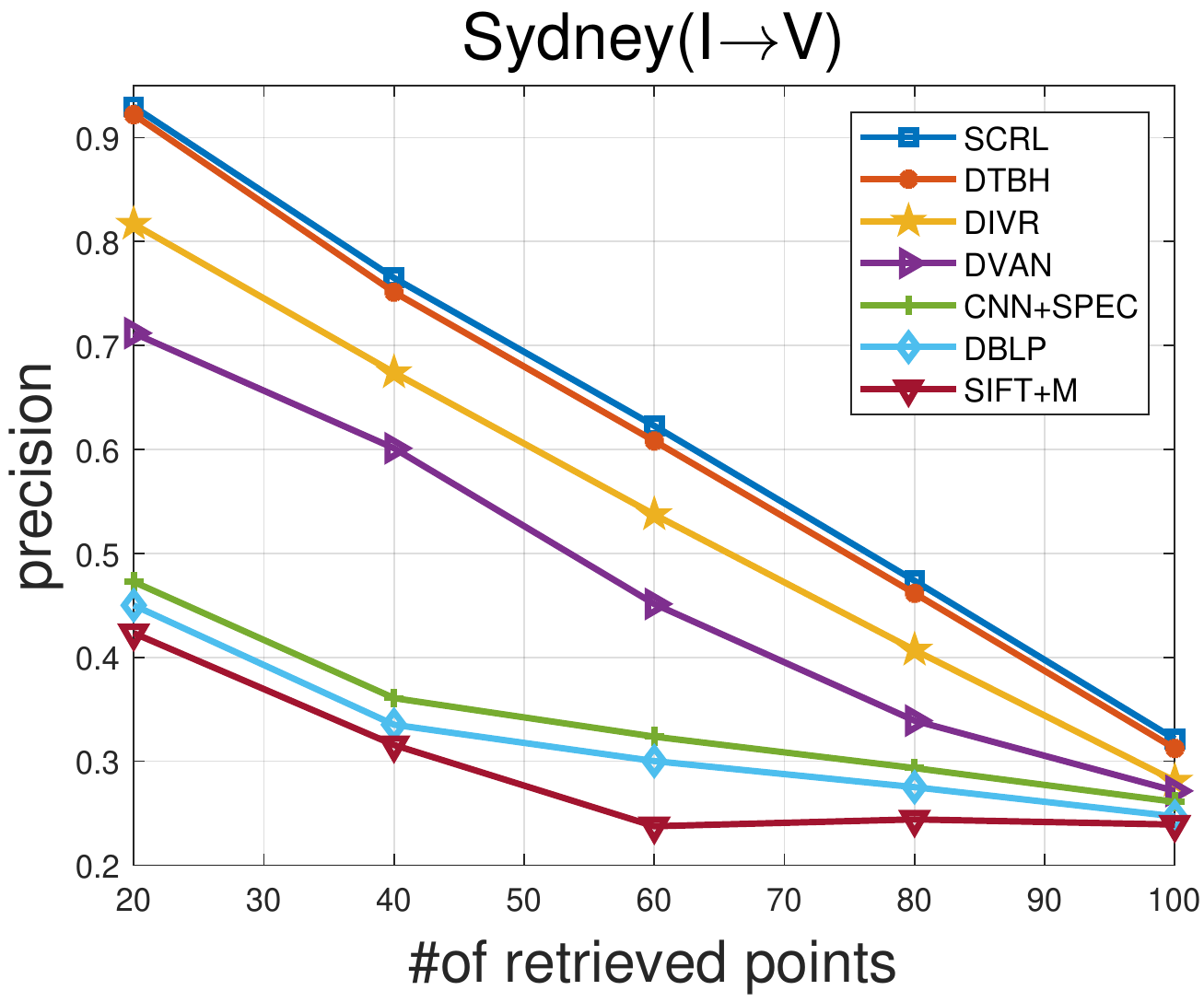}
\renewcommand{\figurename}{Fig.}
\vspace{-8 mm}
\end{center}
    \caption{\small{The precision curves with different retrieved samples by the I$\rightarrow$V protocol on the Sydney IV dataset.}}
\label{fig:5}
\end{figure}

\begin{figure}[tp]
\begin{center}
\includegraphics[width=\linewidth]{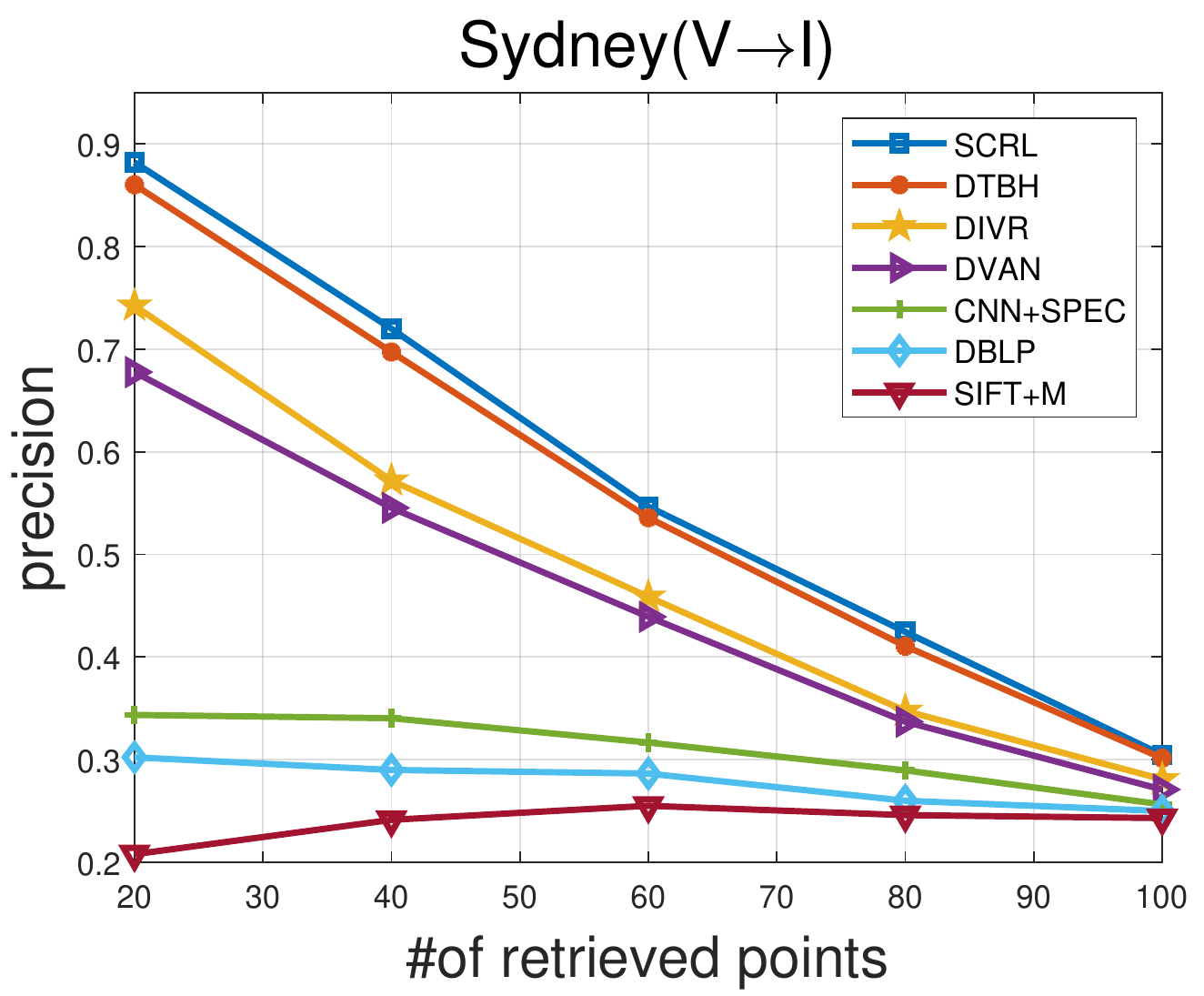}
\renewcommand{\figurename}{Fig.}
\vspace{-8 mm}
\end{center}
    \caption{\small{The precision curves with different retrieved samples by the V$\rightarrow$I protocol on the Sydney IV dataset.}}
\label{fig:6}
\end{figure}

\subsubsection{\textbf{Non-paired Inter-Modality Consistency}} The non-paired inter-modality consistency aims to learn the relationship between two representations in different modalities to keep the semantic consistency across the two modalities. The specific comparison situation corresponds to the results of SCRL-Inter method and the proposed SCRL method in TABLE \ref{tab:2}-\ref{tab:4}. From the results, we can notice the retrieval performance is significantly improved when the intra-modality consistency is considered. This is because the intra-modality consistency loss contributes to narrowing the distance between two representations from different modalities when they describe the same semantic concept.

\subsubsection{\textbf{Semantic Discrimination}} The classification loss aims to enhance the semantic discrimination ability of representations so as to retrieve the relevant samples more easily. The specific comparison situation corresponds to the results of SCRL-Class method and the proposed SCRL method in TABLE \ref{tab:2}-\ref{tab:4}. From the comparison results, we can see the scores of all evaluation metrics are lifted when the classification loss is added. This proves the classification loss contributes to enhancing the semantic discrimination ability of representations.

\subsubsection{\textbf{Long Range Correlation}} The dilation convolution is adopted to capture the long range correlation of each voice sample. The concrete comparison situation corresponds to the results of SCRL-Dilation method and the proposed SCRL method in TABLE \ref{tab:2}-\ref{tab:4}. The results show the retrieval performance is slightly improved, which indicates the dilation convolution can indeed capture the long range correlation relationship within each voice sample by increasing the receptive field.

\begin{table}
  \centering
\small
  \caption{THE COMPARISON RESULTS BETWEEN THE PROPOSED SCRL METHOD AND OTHER METHIDS ON THE UCM IV DATASET.}
  \begin{threeparttable}
 \setlength{\tabcolsep}{1.5mm}{	
    \begin{tabular}{ccccccc}\toprule
     Protocols                        &Methods         & mAP            & $\bf P$@{\textbf 1} &$\bf P$@{\textbf 5} &$\bf P$@{\textbf 10} \\
    \hline
    \multirow{13}*{I$\rightarrow$V}      &SIFT+M \cite{Zhao2018SIFT}             &8.55     &4.56     &4.65     &4.56\\
                                         &DBLP \cite{DBLP:conf/nips/HarwathTG16} &25.48     &24.18     &23.87     &23.24\\
                                         &CNN+SPEC \cite{8237335}                &26.25     &29.50     &25.52     &23.65\\
                                         &DVAN \cite{8486338}                    &36.79     &32.37     &33.29     &33.74\\
                                         &CMIR-NET \cite{chaudhuri2020cmir-net}  &45.82     &52.92     &49.74     &43.38\\
                                         &DIVR \cite{9044618}                    &50.94     &59.34     &54.17     &50.12\\
                                         &DTBH \cite{chen2019a}                  &64.24     &73.10     &69.69     &65.63\\
         \cline{2-6}
                                         &SCRL-Pair                              &35.25     &35.24     &35.48     &35.50\\
                                         &SCRL-Intra                             &52.64     &59.90     &57.10     &54.19\\
                                         &SCRL-Inter                             &53.62     &61.95     &56.33     &52.17\\
                                         &SCRL-Class                             &60.54     &68.81     &64.86     &59.62\\
                                         &SCRL-Dilation                          &65.46     &72.57     &70.43     &67.86\\
                                         &\bf{SCRL}      &\textbf{67.97}    &\textbf{77.38}      &\textbf{75.29}  &\textbf{72.26}\\
    \hline
    \multirow{13}*{V$\rightarrow$I}      &SIFT+M \cite{Zhao2018SIFT}             &6.66     &3.58     &4.41     &4.68\\
                                         &DBLP \cite{DBLP:conf/nips/HarwathTG16} &19.33     &17.12     &17.62     &16.31\\
                                         &CNN+SPEC \cite{8237335}                &21.79     &19.42     &19.86     &19.23\\
                                         &DVAN \cite{8486338}                    &32.28     &32.37     &33.91     &34.34\\
                                         &CMIR-NET \cite{chaudhuri2020cmir-net}  &40.37     &46.74     &43.75     &39.62\\
                                         &DIVR \cite{9044618}                    &45.34     &52.23     &48.76     &44.98\\
                                         &DTBH \cite{chen2019a}                  &60.13     &70.26     &66.63     &61.73\\
       \cline{2-6}
                                         &SCRL-Pair                              &34.69     &39.76     &37.38     &36.76\\
                                         &SCRL-Intra                             &50.74     &56.67     &52.86     &48.40 \\
                                         &SCRL-Inter                             &49.31     &56.86     &53.86     &51.90\\
                                         &SCRL-Class                             &59.82     &66.90     &59.05     &55.76\\
                                         &SCRL-Dilation                          &66.59     &72.62     &68.52     &64.31\\
                                         &\bf{SCRL}      &\textbf{66.83}    &\textbf{84.05}      &\textbf{77.14}  &\textbf{74.07} \\
\bottomrule
    \end{tabular}}%
  \label{tab:3}%
\end{threeparttable}
\end{table}%

\begin{figure*}[tp]
\begin{center}
\includegraphics[width=\linewidth]{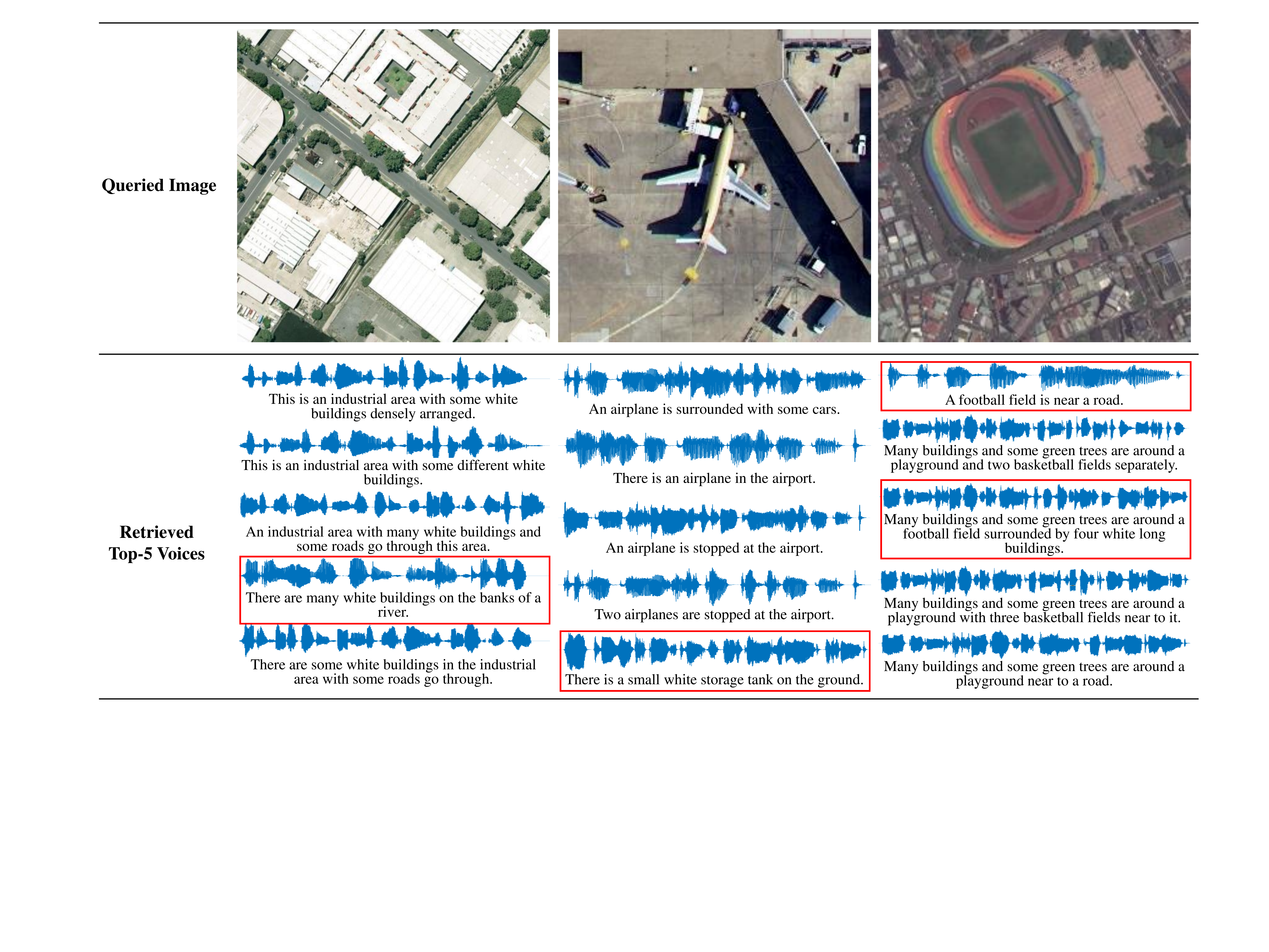}
\renewcommand{\figurename}{Fig.}
\vspace{-5 mm}
\end{center}
    \caption{\small{Some examples of the I$\rightarrow$V retrieval results by the proposed SCRL method on Sydney, UCM and RSICD IV datasets. The result examples on the three datasets correspond to the first column, second column and third column, respectively. The wrong retrieval results are marked with red boxes.}}
\label{fig:7}
\end{figure*}

\subsection{Results and Analysis}
The experimental results and the corresponding analysis on three challenging RS image-voice datasets are given as follows.
\subsubsection{\textbf{Results on Sydney IV Dataset}}
\begin{figure}[tp]
\begin{center}
\includegraphics[width=\linewidth]{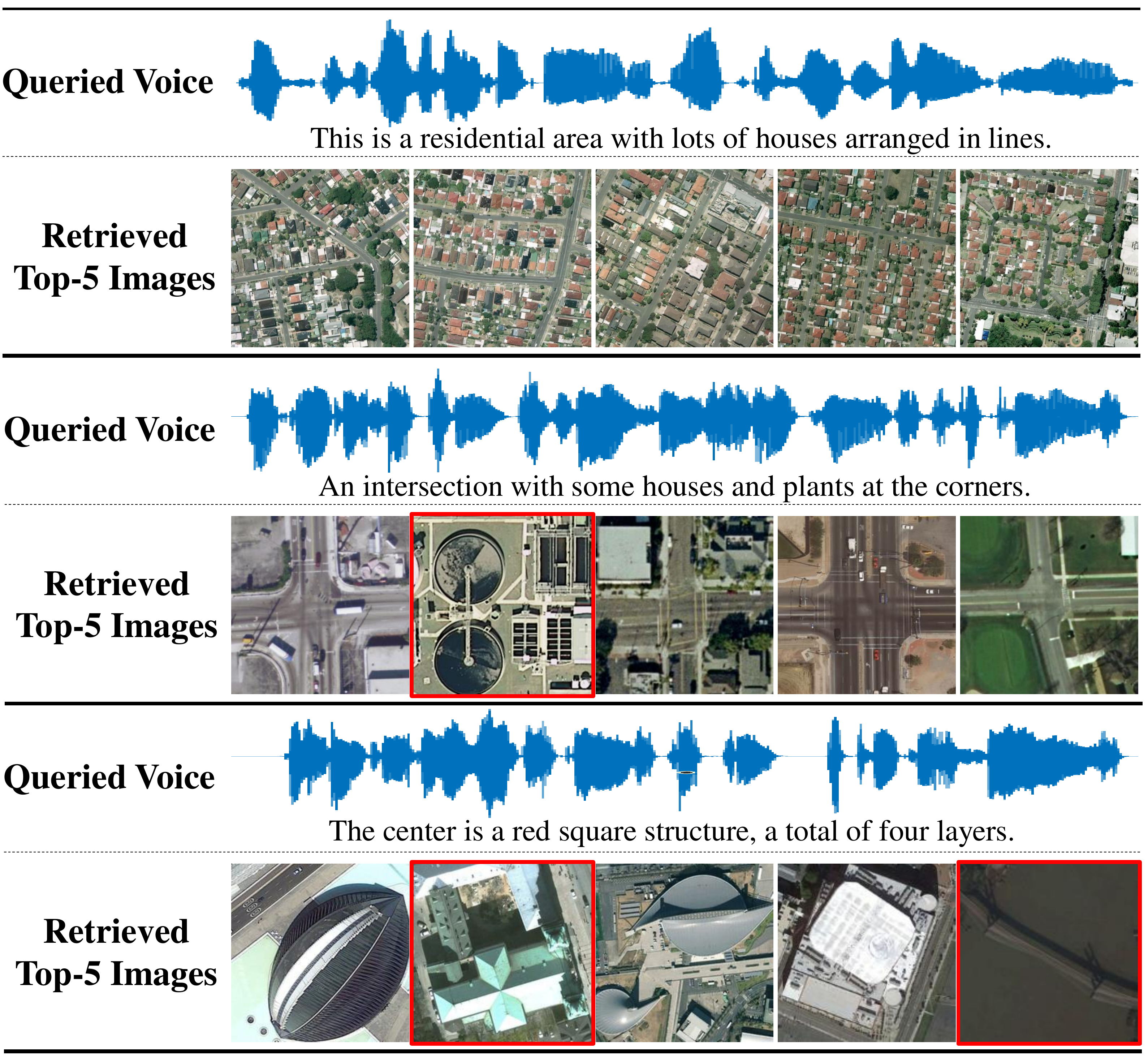}
\renewcommand{\figurename}{Fig.}
\vspace{-5 mm}
\end{center}
    \caption{\small{Some examples of the V$\rightarrow$I retrieval results by the proposed SCRL method on Sydney, UCM and RSICD IV datasets. The result examples on the three datasets correspond to the first row, second row and third row, respectively. The wrong retrieval results are marked with red boxes.}}
\label{fig:8}
\end{figure}

TABLE \ref{tab:2} shows the comparison results between the proposed SCRL method and other compared methods on the Sydney IV dataset. Fig \ref{fig:5} and Fig \ref{fig:6} show the precision curves with different retrieved samples by the I$\rightarrow$V and V$\rightarrow$I protocol, respectively. By observing the results in TABLE \ref{tab:2}, Fig \ref{fig:5} and Fig \ref{fig:6}, we can find: 1) the proposed SCRL method achieves the highest value in terms of most evaluation metrics. 2) Fig \ref{fig:5} and Fig \ref{fig:6} shows the proposed SCRL method outperforms other comparison methods at all returned neighbors. 3) As for the I$\rightarrow$V protocol, the proposed SCRL method improves the mAP value from SIFT+M (31.67\%), DBLP (44.38\%), CNN+SPEC (46.67\%), DVAN (71.77\%), CMIR-NET (78.44\%), DIVR (81.35\%), DTBH (92.45\%) to 94.24\%. Meanwhile, for the V$\rightarrow$I protocol, the proposed SCRL method improves the mAP value from SIFT+M (26.50\%), DBLP (34.87\%), CNN+SPEC (35.72\%), DVAN (63.88\%), CMIR-NET (71.28\%), DIVR (75.97\%), DTBH (87.49\%) to 93.53\%. The improvements demonstrate that the modeling for the pairwise, intra-modality, and non-paired inter-modality relationships contributes to learning semantics-consistency representations across the two modalities, thereby facilitating cross-modal retrieval. In addition, the first column of Fig. \ref{fig:7} shows an example of the top five retrieved results by the proposed SCRL method with the I$\rightarrow$V protocol. The first row of Fig. \ref{fig:8} shows an example of the top five retrieved results by the proposed SCRL method with the V$\rightarrow$I protocol. As shown in the two figures, the proposed SCRL method can retrieve the relevant samples effectively, which further proves the effectiveness for modeling the semantics-consistent relationships comprehensively.

\begin{figure}[tp]
\begin{center}
\includegraphics[width=\linewidth]{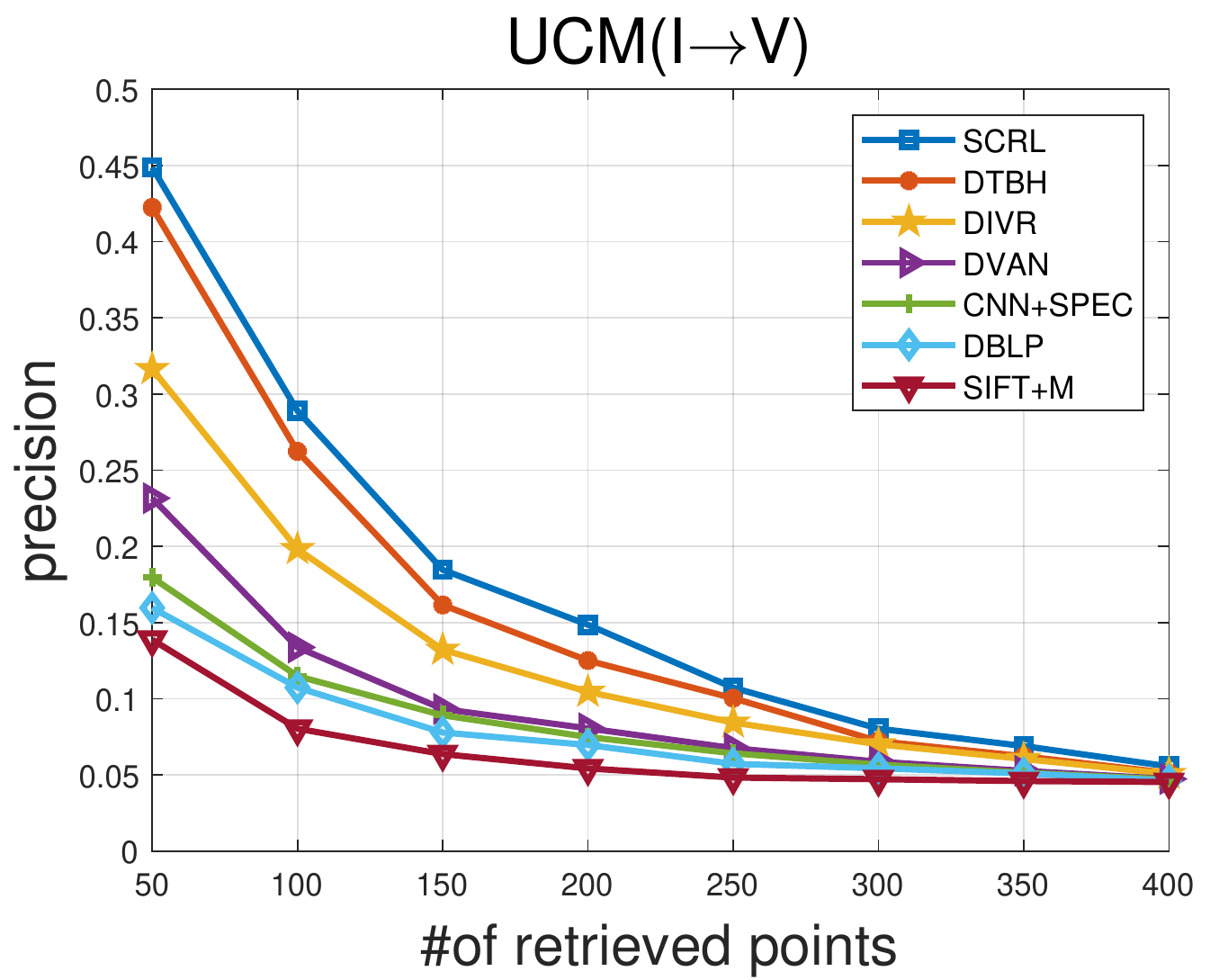}
\renewcommand{\figurename}{Fig.}
\vspace{-8 mm}
\end{center}
    \caption{\small{The precision curves with different retrieved samples by the I$\rightarrow$V protocol on the UCM IV dataset.}}
\label{fig:9}
\end{figure}

\begin{figure}[tp]
\begin{center}
\includegraphics[width=\linewidth]{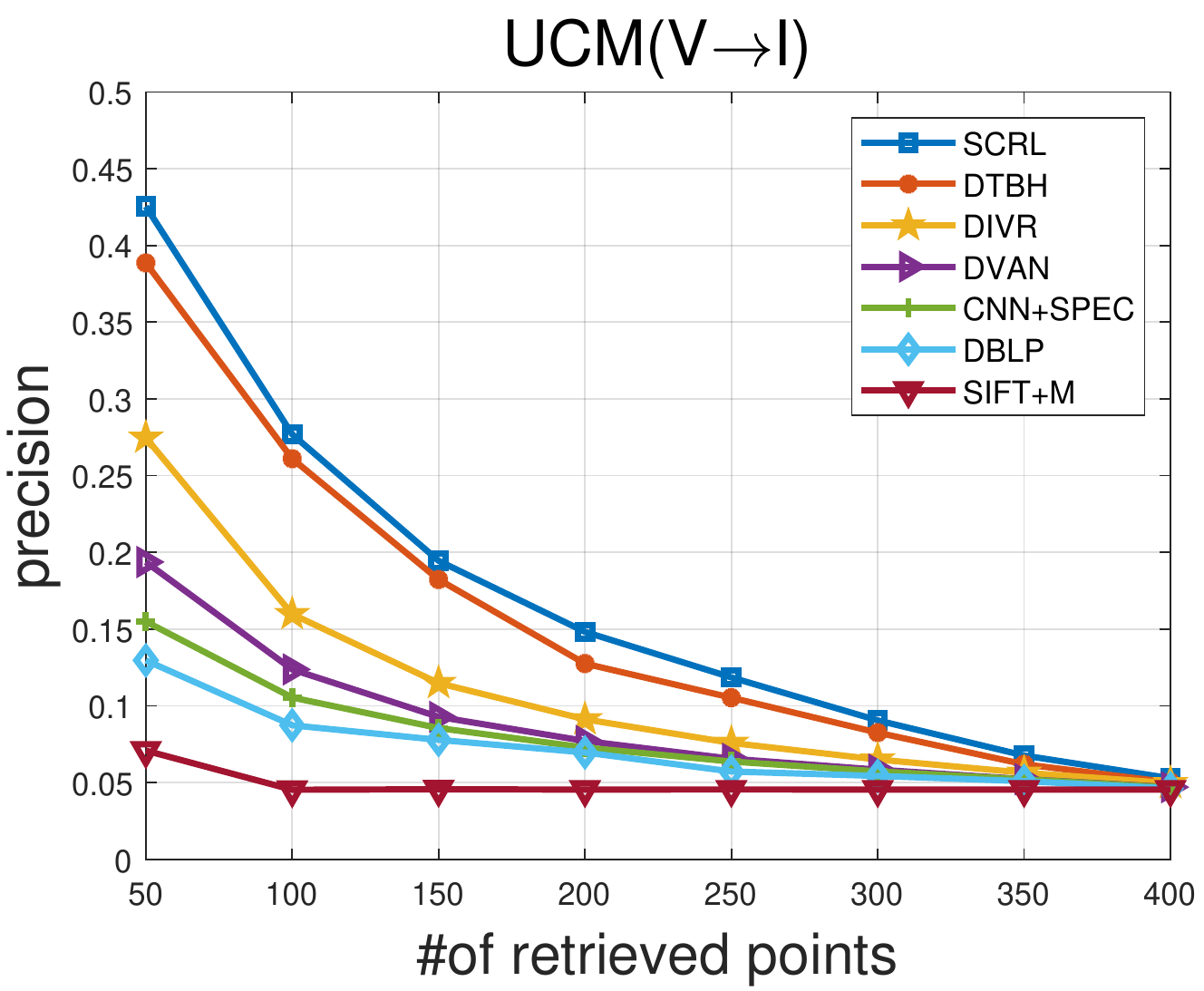}
\renewcommand{\figurename}{Fig.}
\vspace{-8 mm}
\end{center}
    \caption{\small{The precision curves with different retrieved samples by the V$\rightarrow$I protocol on the UCM IV dataset.}}
\label{fig:10}
\end{figure}

\subsubsection{\textbf{Results on UCM IV Dataset}} TABLE \ref{tab:3} shows the comparison results between the proposed SCRL method and other compared methods on the UCM IV Dataset. Fig \ref{fig:9} and Fig \ref{fig:10} show the precision curves with different retrieved samples by the I$\rightarrow$V and V$\rightarrow$I protocol, respectively. Similar experimental results can be seen on the Sydney IV dataset. For instance, for the I$\rightarrow$V protocol, the proposed SCRL method improves the mAP value from SIFT+M (31.67\%), DBLP (44.38\%), CNN+SPEC (46.67\%), DVAN (71.77\%), CMIR-NET (78.44\%), DIVR (81.35\%), DTBH (92.45\%) to 67.97\%. Meanwhile, for the V$\rightarrow$I protocol, the proposed SCRL method improves the mAP value from SIFT+M (26.50\%), DBLP (34.87\%), CNN+SPEC (35.72\%), DVAN (63.88\%), CMIR-NET (71.28\%), DIVR (75.97\%), DTBH (87.49\%) to 66.83\%. This is because the exploration for the semantics-consistent representation space can effectively promote the model to narrow the heterogeneous semantic gap across the two modalities. In addition, the second column of Fig. \ref{fig:7} shows an example of the top five retrieved results by the proposed SCRL method with the I$\rightarrow$V protocol. The second row of Fig. \ref{fig:8} shows an example of the top five retrieved results by the proposed SCRL method with the V$\rightarrow$I protocol. The results show that the proposed SCRL method can effectively retrieve relevant samples, which further proves the effectiveness of considering the pairwise, intra-modality, and non-paired inter-modality relationships comprehensively.

\subsubsection{\textbf{Results on RSICD IV Dataset}} 
\begin{figure}[tp]
\begin{center}
\includegraphics[width=\linewidth]{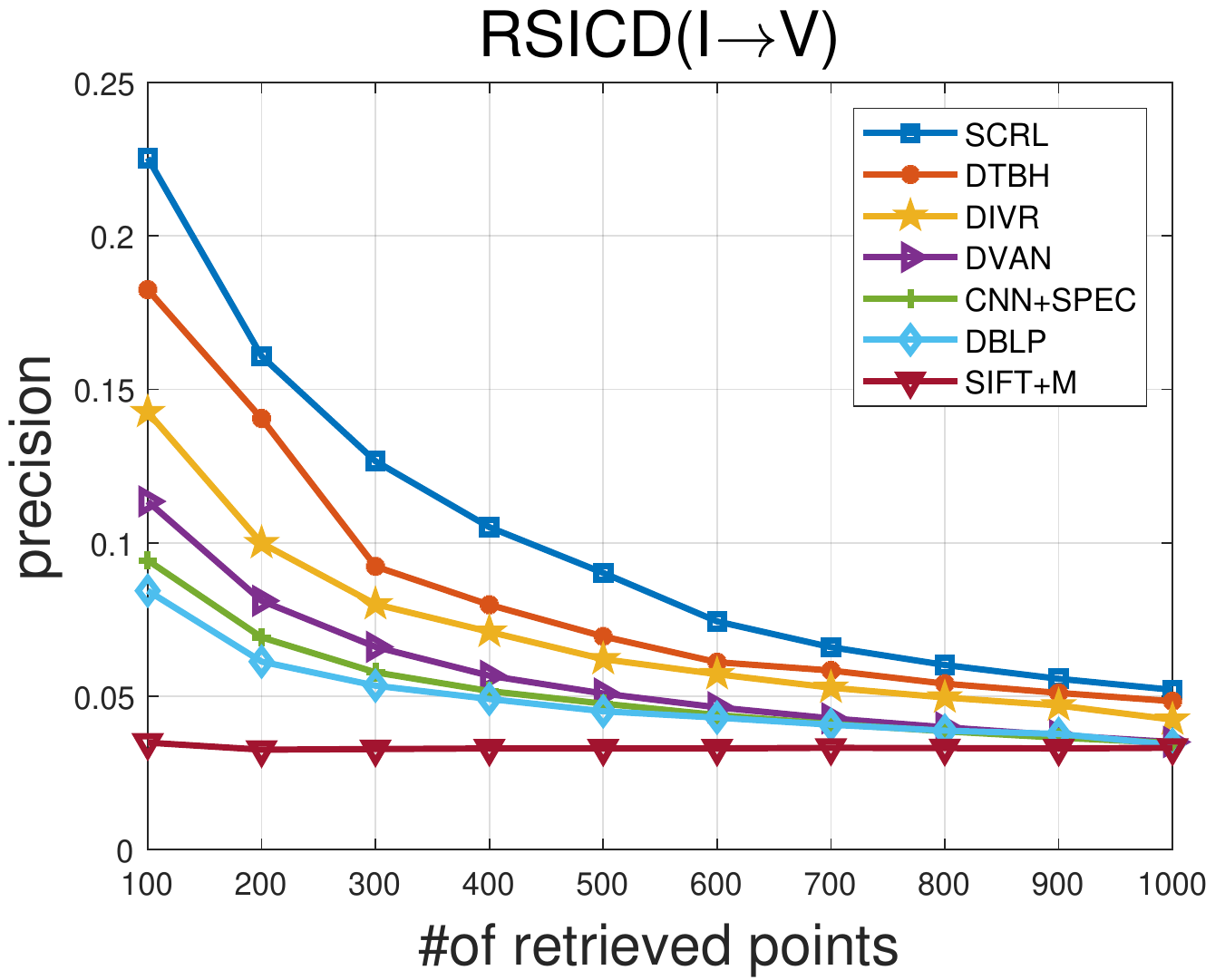}
\renewcommand{\figurename}{Fig.}
\vspace{-8 mm}
\end{center}
    \caption{\small{The precision curves with different retrieved samples by the I$\rightarrow$V protocol on the RSICD IV dataset.}}
\label{fig:11}
\end{figure}

\begin{figure}[tp]
\begin{center}
\includegraphics[width=\linewidth]{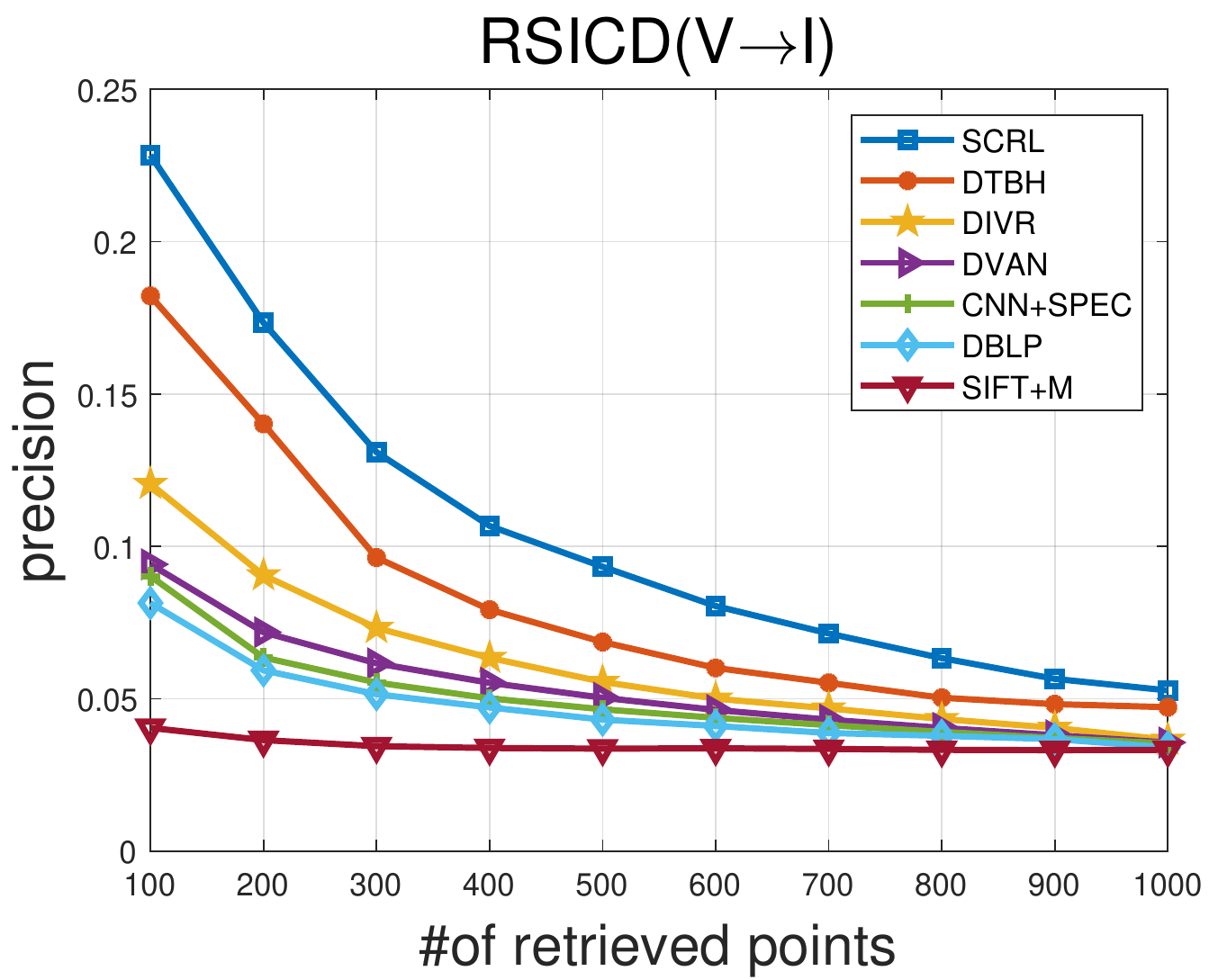}
\renewcommand{\figurename}{Fig.}
\vspace{-8 mm}
\end{center}
    \caption{\small{The precision curves with different retrieved samples by the V$\rightarrow$I protocol on the RSICD IV dataset.}}
\label{fig:12}
\end{figure}

TABLE \ref{tab:4} shows the comparison results between the proposed SCRL method and other compared methods on the most challenging RSICD IV Dataset. Fig \ref{fig:11} and Fig \ref{fig:12} show the precision curves with different retrieved samples by the I$\rightarrow$V and V$\rightarrow$I protocol, respectively. Although the RSICD IV dataset is more challenging \cite{9044618}, the proposed SCRL method can improve the retrieval performance to a great extent. For example, the SCRL method lifts the mAP value from 23.46\% to 28.84\% for the I$\rightarrow$V protocol compared with the state-of-the-art DTBH method. Meanwhile, the SCRL method improves the mAP value from 23.46\% (DTBH) to 28.84\% for the V$\rightarrow$I protocol. This demonstrates the high efficiency of the proposed SCRL method for learning semantics-consistent representations across the two modalities, because the semantics-consistent relationships, including pairwise, intra-modality, and non-paired inter-modality relationships, are modeled comprehensively. In addition, some examples of the top five retrieved results by the proposed SCRL method are shown in Fig. \ref{fig:7} (3rd column) and Fig. \ref{fig:8} (3rd row). It can be seen that the proposed SCRL method can retrieve diverse but relevant samples effectively, even on very challenging image-voice dataset, which further verifies the effectiveness of simultaneously modeling various semantics-consistent relationships.

\renewcommand\arraystretch{1.1}
\begin{table}[tp]
  \centering
\small
  \caption{THE COMPARISON RESULTS BETWEEN THE PROPOSED SCRL METHOD AND OTHER METHIDS ON THE RSICD IV DATASET (\%).}
  \begin{threeparttable}
 \setlength{\tabcolsep}{1.5mm}{	
    \begin{tabular}{ccccccc}\toprule
     Protocols                        &Methods         & mAP            & $\bf P$@{\textbf 1} &$\bf P$@{\textbf 5} &$\bf P$@{\textbf 10} \\
    \hline
    \multirow{13}*{I$\rightarrow$V}      &SIFT+M \cite{Zhao2018SIFT}             & 5.04     &6.22     &5.34     &4.50\\
                                         &DBLP \cite{DBLP:conf/nips/HarwathTG16} &12.70     &15.32     &15.21     &14.22\\
                                         &CNN+SPEC \cite{8237335}                &13.24     &16.82     &16.62     &15.69\\
                                         &DVAN \cite{8486338}                    &16.29     &22.49     &22.56     &21.70\\
                                         &CMIR-NET \cite{chaudhuri2020cmir-net}  &17.78     &24.11     &23.52     &22.54\\
                                         &DIVR \cite{9044618}                    &19.62     &25.43     &24.84    &24.20\\
                                         &DTBH \cite{chen2019a}                  &23.46     &27.58     &26.84     &25.49\\
         \cline{2-6}
                                         &SCRL-Pair                              &15.80     &17.09     &16.84     &16.05\\
                                         &SCRL-Intra                             &20.06     &21.78     &21.34     &20.75\\
                                         &SCRL-Inter                             &20.02     &22.46     &22.44     &21.73\\
                                         &SCRL-Class                             &25.92     &29.54     &28.35     &27.00\\
                                         &SCRL-Dilation                          &27.54     &30.88     &29.96     &29.81\\
                                         &\bf{SCRL}                              &\textbf{28.84} &\textbf{31.10} &\textbf{29.81} &\textbf{28.34}\\
    \hline
    \multirow{13}*{V$\rightarrow$I}      &SIFT+M \cite{Zhao2018SIFT}             &4.85     &3.66     &3.60     &3.541\\
                                         &DBLP \cite{DBLP:conf/nips/HarwathTG16} &8.14     &6.21     &6.08     &6.76\\
                                         &CNN+SPEC \cite{8237335}                &9.96     &7.13     &7.00     &7.44\\
                                         &DVAN \cite{8486338}                    &15.71     &16.18     &15.10     &14.76\\
                                         &CMIR-NET \cite{chaudhuri2020cmir-net}  &17.25     &17.94     &16.58     &15.36\\
                                         &DIVR \cite{9044618}                    &18.58     &19.76     &18.31     &17.59\\
                                         &DTBH \cite{chen2019a}                  &22.72     &23.30     &22.48     &21.17\\
       \cline{2-6}
                                         &SCRL-Pair                              &15.35     &18.56     &17.31     &16.35\\
                                         &SCRL-Intra                             &24.52     &30.11     &29.08     &28.08\\
                                         &SCRL-Inter                             &25.21     &30.60     &30.72     &29.64\\
                                         &SCRL-Class                             &28.33     &31.73     &31.57     &30.83\\
                                         &SCRL-Dilation                          &29.55     &32.59     &31.25     &30.08\\
                                         &\bf{SCRL}      &\textbf{31.26}    &\textbf{33.76}      &\textbf{33.01}  &\textbf{32.01}\\
\bottomrule
    \end{tabular}}%
  \label{tab:4}%
\end{threeparttable}
\end{table}%

\section{Conclusions}  \label{conclutions}
In this paper, a semantics-consistent representation learning (SCRL) method is proposed for the task of remote sensing (RS) image-voice retrieval.
The main novelty is that the proposed method takes the pairwise, intra-modality, and non-paired inter-modality relationships into account simultaneously, thereby improving the semantic consistency of the learned representations for the RS image-voice retrieval. Specifically, a consistent representation space is explored with a joint loss function comprehensively considering the three kinds of relationships for the extracted representations. By exploring the consistent representation space, the heterogeneous semantic gap can be mitigated to a great extent and semantics-consistent representations can be learned for the RS image-voice retrieval. Experimental results on Sydney, UCM and RSICD image-voice datasets demonstrate that the proposed joint loss is effective to mine the pairwise, intra-modality, and non-paired inter-modality relationships for better RS image-voice retrieval. In addition, the proposed SCRL method achieves better performance compared with other state-of-the-art methods, manifesting its superiority. In the future, we will explore how to extend the proposed SCRL method to address the large-scale image-audio retrieval cases such as embedding the hashing code learning to our work.

\bibliographystyle{IEEEtran}
\bibliography{reference}

\begin{thebibliography}{10}
\providecommand{\url}[1]{#1}
\csname url@samestyle\endcsname
\providecommand{\newblock}{\relax}
\providecommand{\bibinfo}[2]{#2}
\providecommand{\BIBentrySTDinterwordspacing}{\spaceskip=0pt\relax}
\providecommand{\BIBentryALTinterwordstretchfactor}{4}
\providecommand{\BIBentryALTinterwordspacing}{\spaceskip=\fontdimen2\font plus
\BIBentryALTinterwordstretchfactor\fontdimen3\font minus
  \fontdimen4\font\relax}
\providecommand{\BIBforeignlanguage}[2]{{%
\expandafter\ifx\csname l@#1\endcsname\relax
\typeout{** WARNING: IEEEtran.bst: No hyphenation pattern has been}%
\typeout{** loaded for the language `#1'. Using the pattern for}%
\typeout{** the default language instead.}%
\else
\language=\csname l@#1\endcsname
\fi
#2}}
\providecommand{\BIBdecl}{\relax}
\BIBdecl

\bibitem{8986556}
Y.~{Hua}, L.~{Mou}, and X.~X. {Zhu}, ``Relation network for multilabel aerial
  image classification,'' \emph{IEEE Transactions on Geoscience and Remote
  Sensing}, vol.~58, no.~7, pp. 4558--4572, 2020.

\bibitem{8385104}
Y.~{Li}, Y.~{Zhang}, X.~{Huang}, and J.~{Ma}, ``Learning source-invariant deep
  hashing convolutional neural networks for cross-source remote sensing image
  retrieval,'' \emph{IEEE Transactions on Geoscience and Remote Sensing},
  vol.~56, no.~11, pp. 6521--6536, 2018.

\bibitem{zhang2019hierarchical}
Y.~Zhang, Y.~Yuan, Y.~Feng, and X.~Lu, ``Hierarchical and robust convolutional
  neural network for very high-resolution remote sensing object detection,''
  \emph{IEEE Transactions on Geoscience and Remote Sensing}, vol.~57, no.~8,
  pp. 5535--5548, 2019.

\bibitem{8067633}
Y.~{Li}, Y.~{Zhang}, X.~{Huang}, H.~{Zhu}, and J.~{Ma}, ``Large-scale remote
  sensing image retrieval by deep hashing neural networks,'' \emph{IEEE
  Transactions on Geoscience and Remote Sensing}, vol.~56, no.~2, pp. 950--965,
  2018.

\bibitem{demir2016hashing-based}
B.~Demir and L.~Bruzzone, ``Hashing-based scalable remote sensing image search
  and retrieval in large archives,'' \emph{IEEE Transactions on Geoscience and
  Remote Sensing}, vol.~54, no.~2, pp. 892--904, 2016.

\bibitem{Chen2020Supervised}
Y.~Chen and X.~Lu, ``Supervised deep hashing with a joint deep network,''
  \emph{Pattern Recognition}, vol. 105, p. 107368, 2020.

\bibitem{li2018large-scale}
Y.~Li, Y.~Zhang, X.~Huang, H.~Zhu, and J.~Ma, ``Large-scale remote sensing
  image retrieval by deep hashing neural networks,'' \emph{IEEE Transactions on
  Geoscience and Remote Sensing}, vol.~56, no.~2, pp. 950--965, 2018.

\bibitem{chaudhuri2020cmir-net}
U.~Chaudhuri, B.~Banerjee, A.~Bhattacharya, and M.~Datcu, ``Cmir-net : A deep
  learning based model for cross-modal retrieval in remote sensing,''
  \emph{Pattern Recognition Letters}, vol. 131, pp. 456--462, 2020.

\bibitem{shi2017can}
Z.~Shi and Z.~Zou, ``Can a machine generate humanlike language descriptions for
  a remote sensing image?'' \emph{IEEE Transactions on Geoscience and Remote
  Sensing}, vol.~55, no.~6, pp. 3623--3634, 2017.

\bibitem{8954885}
Z.~{Shao}, W.~{Zhou}, X.~{Deng}, M.~{Zhang}, and Q.~{Cheng}, ``Multilabel
  remote sensing image retrieval based on fully convolutional network,''
  \emph{IEEE Journal of Selected Topics in Applied Earth Observations and
  Remote Sensing}, vol.~13, pp. 318--328, 2020.

\bibitem{8486338}
G.~{Mao}, Y.~{Yuan}, and L.~{Xiaoqiang}, ``Deep cross-modal retrieval for
  remote sensing image and audio,'' in \emph{2018 10th IAPR Workshop on Pattern
  Recognition in Remote Sensing}, 2018, pp. 1--7.

\bibitem{8896932}
M.~{Guo}, C.~{Zhou}, and J.~{Liu}, ``Jointly learning of visual and auditory: A
  new approach for rs image and audio cross-modal retrieval,'' \emph{IEEE
  Journal of Selected Topics in Applied Earth Observations and Remote Sensing},
  vol.~12, no.~11, pp. 4644--4654, 2019.

\bibitem{chen2019a}
Y.~Chen and X.~Lu, ``A deep hashing technique for remote sensing image-sound
  retrieval,'' \emph{Remote Sensing}, vol.~12, no.~1, p.~84, 2019.

\bibitem{9044618}
Y.~{Chen}, X.~{Lu}, and S.~{Wang}, ``Deep cross-modal image-voice retrieval in
  remote sensing,'' \emph{IEEE Transactions on Geoscience and Remote Sensing},
  vol.~58, no.~10, pp. 7049--7061, 2020.

\bibitem{shao2020multilabel}
Z.~Shao, W.~Zhou, X.~Deng, M.~Zhang, and Q.~Cheng, ``Multilabel remote sensing
  image retrieval based on fully convolutional network,'' \emph{IEEE Journal of
  Selected Topics in Applied Earth Observations and Remote Sensing}, vol.~13,
  pp. 318--328, 2020.

\bibitem{9046296}
M.~{Lin}, R.~{Ji}, S.~{Chen}, X.~{Sun}, and C.~{Lin}, ``Similarity-preserving
  linkage hashing for online image retrieval,'' \emph{IEEE Transactions on
  Image Processing}, vol.~29, pp. 5289--5300, 2020.

\bibitem{Simonyan2014Very}
K.~Simonyan and A.~Zisserman, ``Very deep convolutional networks for
  large-scale image recognition,'' \emph{Computer Science}, 2014.

\bibitem{8534807}
R.~{Hidayat}, A.~{Bejo}, S.~{Sumaryono}, and A.~{Winursito}, ``Denoising speech
  for mfcc feature extraction using wavelet transformation in speech
  recognition system,'' in \emph{2018 10th International Conference on
  Information Technology and Electrical Engineering}, 2018, pp. 280--284.

\bibitem{luo2008indexing}
B.~Luo, J.~Aujol, Y.~Gousseau, and S.~Ladjal, ``Indexing of satellite images
  with different resolutions by wavelet features,'' \emph{IEEE Transactions on
  Image Processing}, vol.~17, no.~8, pp. 1465--1472, 2008.

\bibitem{yang2013geographic}
Y.~Yang and S.~Newsam, ``Geographic image retrieval using local invariant
  features,'' \emph{IEEE Transactions on Geoscience and Remote Sensing},
  vol.~51, no.~2, pp. 818--832, 2013.

\bibitem{aptoula2014remote}
E.~Aptoula, ``Remote sensing image retrieval with global morphological texture
  descriptors,'' \emph{IEEE Transactions on Geoscience and Remote Sensing},
  vol.~52, no.~5, pp. 3023--3034, 2014.

\bibitem{ning2020audio}
H.~Ning, X.~Zheng, Y.~Yuan, and X.~Lu, ``Audio description from image by modal
  translation network,'' \emph{Neurocomputing}, 2020.

\bibitem{zhao2019weather}
B.~Zhao, L.~Hua, X.~Li, X.~Lu, and Z.~Wang, ``Weather recognition via
  classification labels and weather-cue maps,'' \emph{Pattern Recognition},
  vol.~95, pp. 272--284, 2019.

\bibitem{yuan2018remote}
Y.~Yuan, J.~Fang, X.~Lu, and Y.~Feng, ``Remote sensing image scene
  classification using rearranged local features,'' \emph{IEEE Transactions on
  Geoscience and Remote Sensing}, vol.~57, no.~3, pp. 1779--1792, 2018.

\bibitem{zhao2019cam}
B.~Zhao, X.~Li, and X.~Lu, ``Cam-rnn: Co-attention model based rnn for video
  captioning,'' \emph{IEEE Transactions on Image Processing}, vol.~28, no.~11,
  pp. 5552--5565, 2019.

\bibitem{yuan2019spatial}
Y.~Yuan, J.~Fang, X.~Lu, and Y.~Feng, ``Spatial structure preserving feature
  pyramid network for semantic image segmentation,'' \emph{ACM Transactions on
  Multimedia Computing, Communications, and Applications}, vol.~15, no.~3, pp.
  73:1--73:19, 2019.

\bibitem{tang2017two-stage}
X.~Tang, L.~Jiao, W.~J. Emery, F.~Liu, and D.~Zhang, ``Two-stage reranking for
  remote sensing image retrieval,'' \emph{IEEE Transactions on Geoscience and
  Remote Sensing}, vol.~55, no.~10, pp. 5798--5817, 2017.

\bibitem{ye2018remote}
F.~Ye, H.~Xiao, X.~Zhao, M.~Dong, W.~Luo, and W.~Min, ``Remote sensing image
  retrieval using convolutional neural network features and weighted
  distance,'' \emph{IEEE Geoscience and Remote Sensing Letters}, vol.~15,
  no.~10, pp. 1535--1539, 2018.

\bibitem{chaudhuri2017multilabel}
B.~Chaudhuri, B.~Demir, S.~Chaudhuri, and L.~Bruzzone, ``Multilabel remote
  sensing image retrieval using a semisupervised graph-theoretic method,''
  \emph{IEEE Transactions on Geoscience and Remote Sensing}, vol.~56, no.~2,
  pp. 1144--1158, 2017.

\bibitem{kang2020graph}
J.~Kang, R.~Fernandez-Beltran, D.~Hong, J.~Chanussot, and A.~Plaza, ``Graph
  relation network: Modeling relations between scenes for multilabel
  remote-sensing image classification and retrieval,'' \emph{IEEE Transactions
  on Geoscience and Remote Sensing}, 2020.

\bibitem{liu2020similarity}
Y.~Liu, L.~Ding, C.~Chen, and Y.~Liu, ``Similarity-based unsupervised deep
  transfer learning for remote sensing image retrieval,'' \emph{IEEE
  Transactions on Geoscience and Remote Sensing}, vol.~58, no.~11, pp.
  7872--7889, 2020.

\bibitem{ye2019sar}
F.~Ye, W.~Luo, M.~Dong, H.~He, and W.~Min, ``Sar image retrieval based on
  unsupervised domain adaptation and clustering,'' \emph{IEEE Geoscience and
  Remote Sensing Letters}, vol.~16, no.~9, pp. 1482--1486, 2019.

\bibitem{song2020unified}
J.~Song, T.~He, L.~Gao, X.~Xu, A.~Hanjalic, and H.~T. Shen, ``Unified binary
  generative adversarial network for image retrieval and compression,''
  \emph{International Journal of Computer Vision}, vol. 128, pp.
  2243–--22\,646, 2020.

\bibitem{9044737}
W.~{Xiong}, Z.~{Xiong}, Y.~{Cui}, and Y.~{Lv}, ``A discriminative distillation
  network for cross-source remote sensing image retrieval,'' \emph{IEEE Journal
  of Selected Topics in Applied Earth Observations and Remote Sensing},
  vol.~13, pp. 1234--1247, 2020.

\bibitem{xiong2020deep}
W.~Xiong, Z.~Xiong, Y.~Zhang, Y.~Cui, and X.~Gu, ``A deep cross-modality
  hashing network for sar and optical remote sensing images retrieval,''
  \emph{IEEE Journal of Selected Topics in Applied Earth Observations and
  Remote Sensing}, vol.~13, pp. 5284--5296, 2020.

\bibitem{song2013inter}
J.~Song, Y.~Yang, Y.~Yang, Z.~Huang, and H.~T. Shen, ``Inter-media hashing for
  large-scale retrieval from heterogeneous data sources,'' in \emph{Proceedings
  of the 2013 ACM SIGMOD International Conference on Management of Data}, 2013,
  pp. 785--796.

\bibitem{Noh2017Large}
H.~Noh, A.~Araujo, J.~Sim, T.~Weyand, and B.~Han, ``Large-scale image retrieval
  with attentive deep local features,'' in \emph{2017 IEEE International
  Conference on Computer Vision}, 2017, pp. 3456--3465.

\bibitem{lu2018hierarchical}
X.~Lu, Y.~Chen, and X.~Li, ``Hierarchical recurrent neural hashing for image
  retrieval with hierarchical convolutional features,'' \emph{IEEE Transactions
  on Image Processing}, vol.~27, no.~1, pp. 106--120, 2018.

\bibitem{2018Large}
Y.~Cui, Y.~Song, C.~Sun, A.~Howard, and S.~Belongie, ``Large scale fine-grained
  categorization and domain-specific transfer learning,'' in \emph{2018
  IEEE/CVF Conference on Computer Vision and Pattern Recognition}, 2018, pp.
  4109--4118.

\bibitem{chen2018end-to-end}
Z.~Chen, T.~Zhang, and C.~Ouyang, ``End-to-end airplane detection using
  transfer learning in remote sensing images,'' \emph{Remote Sensing}, vol.~10,
  no.~1, p. 139, 2018.

\bibitem{gatys2016image}
L.~A. Gatys, A.~S. Ecker, and M.~Bethge, ``Image style transfer using
  convolutional neural networks,'' in \emph{Proceedings of the IEEE Conference
  on Computer Vision and Pattern Recognition}, 2016, pp. 2414--2423.

\bibitem{chowdhury2020fusing}
A.~Chowdhury and A.~Ross, ``Fusing mfcc and lpc features using 1d triplet cnn
  for speaker recognition in severely degraded audio signals,'' \emph{IEEE
  Transactions on Information Forensics and Security}, vol.~15, pp. 1616--1629,
  2020.

\bibitem{Chen2020Deep}
Y.~Chen and X.~Lu, ``Deep category-level and regularized hashing with global
  semantic similarity learning,'' \emph{IEEE Transactions on Cybernetics},
  2020.

\bibitem{8985543}
W.~{Xiong}, Y.~{Lv}, X.~{Zhang}, and Y.~{Cui}, ``Learning to translate for
  cross-source remote sensing image retrieval,'' \emph{IEEE Transactions on
  Geoscience and Remote Sensing}, vol.~58, no.~7, pp. 4860--4874, 2020.

\bibitem{chen2020deep_PAIR}
Y.~Chen and X.~Lu, ``Deep discrete hashing with pairwise correlation
  learning,'' \emph{Neurocomputing}, vol. 385, pp. 111--121, 2020.

\bibitem{yuan2019bio-inspired}
Y.~Yuan, H.~Ning, and X.~Lu, ``Bio-inspired representation learning for visual
  attention prediction,'' \emph{IEEE Transactions on Cybernetics}, pp. 1--14,
  2019.

\bibitem{yuan20193g}
A.~Yuan, X.~Li, and X.~Lu, ``3g structure for image caption generation,''
  \emph{Neurocomputing}, vol. 330, pp. 17--28, 2019.

\bibitem{7532802}
T.~{Zhi}, L.~{Duan}, Y.~{Wang}, and T.~{Huang}, ``Two-stage pooling of deep
  convolutional features for image retrieval,'' in \emph{2016 IEEE
  International Conference on Image Processing}, 2016, pp. 2465--2469.

\bibitem{Hu2019Deep}
D.~Hu, F.~Nie, and X.~Li, ``Deep binary reconstruction for cross-modal
  hashing,'' \emph{IEEE Transactions on Multimedia}, vol.~21, no.~4, pp.
  973--985, 2019.

\bibitem{Shao2019Cloud}
Z.~Shao, Y.~Pan, C.~Diao, and J.~Cai, ``Cloud detection in remote sensing
  images based on multiscale features-convolutional neural network,''
  \emph{IEEE Transactions on Geoscience and Remote Sensing}, vol.~57, no.~6,
  pp. 4062--4076, 2019.

\bibitem{khurshid2020a}
N.~Khurshid, M.~Tharani, M.~Taj, and F.~Z. Qureshi, ``A residual-dyad encoder
  discriminator network for remote sensing image matching,'' \emph{IEEE
  Transactions on Geoscience and Remote Sensing}, vol.~58, no.~3, pp.
  2001--2014, 2020.

\bibitem{li2017partial}
P.~Li and P.~Ren, ``Partial randomness hashing for large-scale remote sensing
  image retrieval,'' \emph{IEEE Geoscience and Remote Sensing Letters},
  vol.~14, no.~3, pp. 464--468, 2017.

\bibitem{Zhao2018SIFT}
X.~Zhao, Z.~Li, and J.~Yi, ``Sift feature-based second-order image hash
  retrieval approach,'' \emph{Journal of Software}, vol.~13, no.~1, pp.
  103--116, 2018.

\bibitem{DBLP:conf/nips/HarwathTG16}
D.~F. Harwath, A.~Torralba, and J.~R. Glass, ``Unsupervised learning of spoken
  language with visual context,'' in \emph{Advances in Neural Information
  Processing Systems}, 2016, pp. 1858--1866.

\bibitem{8237335}
R.~{Arandjelovic} and A.~{Zisserman}, ``Look, listen and learn,'' in \emph{2017
  IEEE International Conference on Computer Vision}, 2017, pp. 609--617.

\bibitem{wang2017adversarial}
B.~Wang, Y.~Yang, X.~Xu, A.~Hanjalic, and H.~T. Shen, ``Adversarial cross-modal
  retrieval,'' in \emph{Proceedings of the 25th ACM International Conference on
  Multimedia}, 2017, pp. 154--162.

\end{thebibliography}

\end{document}